\documentclass[11pt,a4paper]{article}
\usepackage{jcappub}

\usepackage{bm}
\usepackage{amsfonts}
\usepackage{subfigure}

\newcommand{\be}{\begin{equation}}
\newcommand{\ee}{\end{equation}}
\newcommand{\bea}{\begin{eqnarray}}
\newcommand{\eea}{\end{eqnarray}}

\newcommand{\di}{\text{d}}

\begin{document}



\title{Measuring neutrino masses with a future galaxy survey}

\author[a]{Jan Hamann}
\author[a]{Steen Hannestad}
\author[b]{Yvonne Y.Y. Wong}

\affiliation[a]{Department of Physics and Astronomy\\
 University of Aarhus, DK-8000 Aarhus C, Denmark}

\affiliation[b]{Institut f\"ur Theoretische Teilchenphysik und Kosmologie \\ RWTH Aachen,
D-52056 Aachen, Germany}

\emailAdd{hamann@phys.au.dk, sth@phys.au.dk, ywong@physik.rwth-aachen.de}

\abstract{We perform a detailed forecast on how well a {\sc Euclid}-like photometric galaxy and cosmic shear survey will be able to constrain the absolute neutrino mass scale. Adopting conservative assumptions about the survey specifications and assuming complete ignorance of the galaxy bias, we estimate that the minimum mass sum of $\sum m_\nu \simeq 0.06$~eV in the normal hierarchy  can be detected at 1.5$\sigma$ to 2.5$\sigma$ significance, depending on the model complexity, using a combination of galaxy and cosmic shear power spectrum measurements in conjunction with CMB temperature and polarisation observations from {\sc Planck}. With better knowledge of the galaxy bias, the significance of the detection could potentially reach~$5.4\sigma$.   Interestingly, neither {\sc Planck}+shear nor {\sc Planck}+galaxy alone can achieve this level of sensitivity;
it is the combined effect of galaxy and cosmic shear power spectrum measurements that  breaks the persistent degeneracies between the neutrino mass, the physical matter density, and the Hubble parameter.  Notwithstanding this remarkable sensitivity to $\sum m_\nu$, {\sc Euclid}-like shear and galaxy data will not be sensitive to the exact mass spectrum of the neutrino sector; no significant bias ($< 1 \sigma$) in the parameter estimation is induced by fitting inaccurate models of the neutrino mass splittings to the mock data, nor does the goodness-of-fit of these models suffer any significant degradation relative to the true one ($\Delta \chi_{\rm eff} ^2< 1$).}

\maketitle

\section{Introduction}

The past decade has seen tremendous advances in the field of precision cosmology.  Measurements of the cosmic microwave background (CMB) temperature and polarisation anisotropies, especially by the WMAP mission~\cite{Komatsu:2010fb}, have yielded information on the structure and content of the universe on length scales up to the current Hubble distance.  Likewise, large-volume galaxy surveys such as the Sloan Digital Sky Survey (SDSS)~\cite{Aihara:2011sj},  which has catalogued the redshifts and angular positions of close to a million galaxies, provide a handle on the distribution of matter on intermediate scales.

The coming decade is likely to be just as exciting. The {\sc Planck} mission~\cite{Planck:2006aa} will measure the CMB anisotropies to an unprecedented precision, and new techniques designed to map out the structures in the low-redshift universe will reach maturity.  Of these new techniques, the most promising is that based on the detection of cosmic shear, i.e.,  the weak gravitational lensing of distant galaxy images by the intervening large-scale structure.  Cosmic shear as a probe of the large-scale matter distribution has
already been demonstrated to yield competitive cosmological constraints with the CFHTLS~\cite{Fu:2007qq} and the COSMOS~\cite{Massey:2007gh} surveys.
Upcoming projects such as the Dark Energy Survey (DES)~\cite{Sanchez:2010zz}, the Large Synoptic Survey Telescope (LSST)~\cite{Abell:2009aa}, and {\sc Euclid}~\cite{Laureijs:2011mu} will take cosmic shear to a new level through an enormous increase in the survey volume.

In this paper, we adopt the technical specifications projected for the photometric redshift survey of  ESA's {\sc Euclid} mission, which has been approved for launch in 2019,
and focus specifically on the cosmological information contained therein.  The {\sc Euclid} photometric survey will be dedicated to the measurement of cosmic shear.
At the same time, such a measurement will also contain the necessary information to enable the use of the galaxy data as a tracer of the underlying (angular) matter distribution.
This dual use of the survey data turns out to be a great advantage for cosmological parameter estimation, especially for the measurement of the absolute neutrino mass scale.  As we shall show, when used in combination, a {\sc Euclid}-like cosmic shear and galaxy distribution measurement has the potential to detect the sum of the neutrino masses $\sum m_\nu$ at better than $5 \sigma$ significance, even if the true $\sum m_\nu$ is the minimum allowed by neutrino oscillation experiments ($ \sim 0.06$~eV).
 This is a significantly better performance than using either shear or galaxy distribution data alone.

The paper is organised as follows.  In section~\ref{sec:observables} we describe the angular power spectrum formalism which we use to quantify both the shear and the galaxy angular clustering two-point statistics.  Section~\ref{sec:errors} discusses the statistical and systematic errors pertinent to the measurements, while
in sections~\ref{sec:mockdatageneration} and~\ref{sec:forecasting} we outline our mock data generation and forecasting procedures, generalising the Markov Chain
Monte Carlo method introduced in reference~\cite{Perotto:2006rj}.
Our results are presented in sections~\ref{sec:results} and \ref{sec:complex} for minimal and non-minimal cosmological models, respectively.
We conclude in section~\ref{sec:conclusions}.  A detailed semi-analytic analysis of the parameter degeneracies in the cosmological probes considered
can be found in the appendix.

\section{{\sc Euclid} observables\label{sec:observables}}

We focus in this work on the photometric galaxy and cosmic shear surveys, which will be one of the main data products from {\sc Euclid}.
For the two observables---the positions of galaxies and the corresponding shear in their images, we quantify their clustering statistics at lowest order
in terms of an angular harmonic power spectrum.    We opt for this measure instead of the more familiar three-dimensional power spectrum
(as in, e.g.,~\cite{Cole:2005sx}), because of the limits imposed by photometry.  The reconstruction of the
three-dimensional power spectrum requires an accurate measurement of the redshift space distortions along the line-of-sight.  While this is well within the capacity of surveys that employ spectroscopy for redshift determination, photometry has too large an uncertainty  to enable this measurement.  The angular power spectrum was recently adopted in the analysis of the SDSS III photometric survey~\cite{Ho:2012vy}.

\subsection{Angular power spectra}

From the two types of observables in our framework, the angular distributions of galaxies~(g) and  shear~(s) on the sky, we can construct  three distinct types of two-point statistics in harmonic space: the shear auto-spectrum $C^{\rm ss}_\ell$, the galaxy auto-spectrum $C^{\rm gg}_\ell$, and the shear-galaxy cross-spectrum $C^{\rm sg}_\ell$, where the subscript $\ell$ refers to the angular multipole.  The shear observable can itself be decomposed into a curl-free $\epsilon$-mode and a divergence-free $\beta$-mode.
However, lensing by scalar metric perturbations at linear order does not produce $\beta$-mode shear.  Henceforth,  the label ``s'' shall denote exclusively $\epsilon$-mode shear.
We also consider the possibility that the shear and the galaxy angular distributions will be measured in several redshift intervals, and adopt the notation $C^{XY}_{\ell,ij}$ to denote the angular auto- or cross-spectrum of the observables $X,Y = {\rm s},{\rm g}$ in the redshift bins $i$ and $j$.

Given a cosmological model, the angular power spectra can be computed from theory according to
\begin{equation}
\label{eq:cl}
  \mathcal{C}_{\ell,ij}^{XY} = 4 \pi \int \di \ln k \;  \mathcal{S}_{\ell,i}^X(k) \, \mathcal{S}_{\ell,j}^Y(k) \, {\mathcal P}_{\mathcal R}(k).
\end{equation}
Here,  $ {\mathcal P}_{\mathcal R}(k)$ is the dimensionless power spectrum of the primordial curvature perturbations~${\mathcal R}(k)$, and
the source functions for shear and galaxies read
\begin{align}
\label{eq:source1}
  \mathcal{S}_{\ell,i}^{\mathrm s} & =  -2 \ \sqrt{\frac{\ell (\ell^2 - 1) (\ell+2)}{4}} \int \di \chi \; j_\ell (k \chi) \, \mathcal{W}_i^{\rm s}(\chi)  \, T_\Psi(k,\eta_0 - \chi), \\
 \label{eq:source2}
  \mathcal{S}_{\ell,i}^{\mathrm g} & =  \int \di \chi \; j_\ell (k \chi) \, \mathcal{W}_i^{\rm g} (\chi) \, T_\delta(k,\eta_0 - \chi) ,
\end{align}
 where the integration variable $\chi$ is the comoving distance, $\eta_0$ the conformal time today, $j_\ell$ are the spherical Bessel functions,%
\footnote{For non-flat universes, $\chi$ should be replaced with the comoving angular diameter distance and $j_\ell$ with the ultra-spherical Bessel function~\cite{Hu:2000ee}.}
 and $\mathcal{T}_{\Psi, \delta}$ are the transfer functions of the metric perturbations%
\footnote{We use the conformal Newtonian gauge whose line element is $\di s^2 = - a^2(\eta) [(1+2 \psi) \di \eta^2 + (1-2 \phi) \gamma_{ij} \di x^i \di x^j]$.
}
 $\Psi \equiv (\psi +\phi)/2$
 and the matter density fluctuation $\delta$ respectively, defined via $\Psi(k,\eta) = T_\Psi(k,\eta) {\mathcal R}(k)$ and
 $\delta(k,\eta) = T_\delta(k,\eta) {\mathcal R}(k)$.  In $\Lambda$CDM-type cosmologies,  the two transfer functions are related to one another at low redshifts
 via the Poisson equation $k^2 T_\Psi (k,\eta) = 4 \pi a^2(\eta) \bar{\rho}_{\rm m}(\eta) T_\delta(k,\eta)$ on subhorizon scales.

The source functions~(\ref{eq:source1}) and~(\ref{eq:source2}) contain the window  functions $\mathcal{W}^X_i$  given by
\begin{align}
  \mathcal{W}_i^{\mathrm g}  ( \chi)& = \int_\chi^\infty \di \chi' \; \frac{\chi-\chi'}{\chi'\chi} \hat{n}_i(\chi'), \\
  \mathcal{W}_i^{\mathrm g}(\chi) & = b(k,\chi)  \hat{n}_i(\chi),
\end{align}
where
\begin{equation}
\label{eq:galredshift}
  \hat{n}_i(\chi) = H(z) \hat{n}_i(z) = H(z) \frac{\di n/\di z(z)}{n_i},
  \end{equation}
encodes the redshift distribution of the source galaxies normalised to the surface density (i.e., number of galaxies per unit solid angle)
\begin{equation}
\label{eq:ni}
 n_i \equiv \int_{\Delta z_i} \di z' \; \di n(z')/\di z'
 \end{equation}
in the redshift bin~$i$, and $\di n(z)/\di z$ is a survey-specific function to be specified in section~\ref{sec:source}.
The galaxy window function $\mathcal{W}_i^{\mathrm g}$ contains an additional dependence on the
galaxy bias $b(k,\chi)$, which relates the galaxy number density fluctuations to the underlying matter density perturbations.

\begin{figure}[t]
\center
\includegraphics[height=.66\textwidth,angle=270]{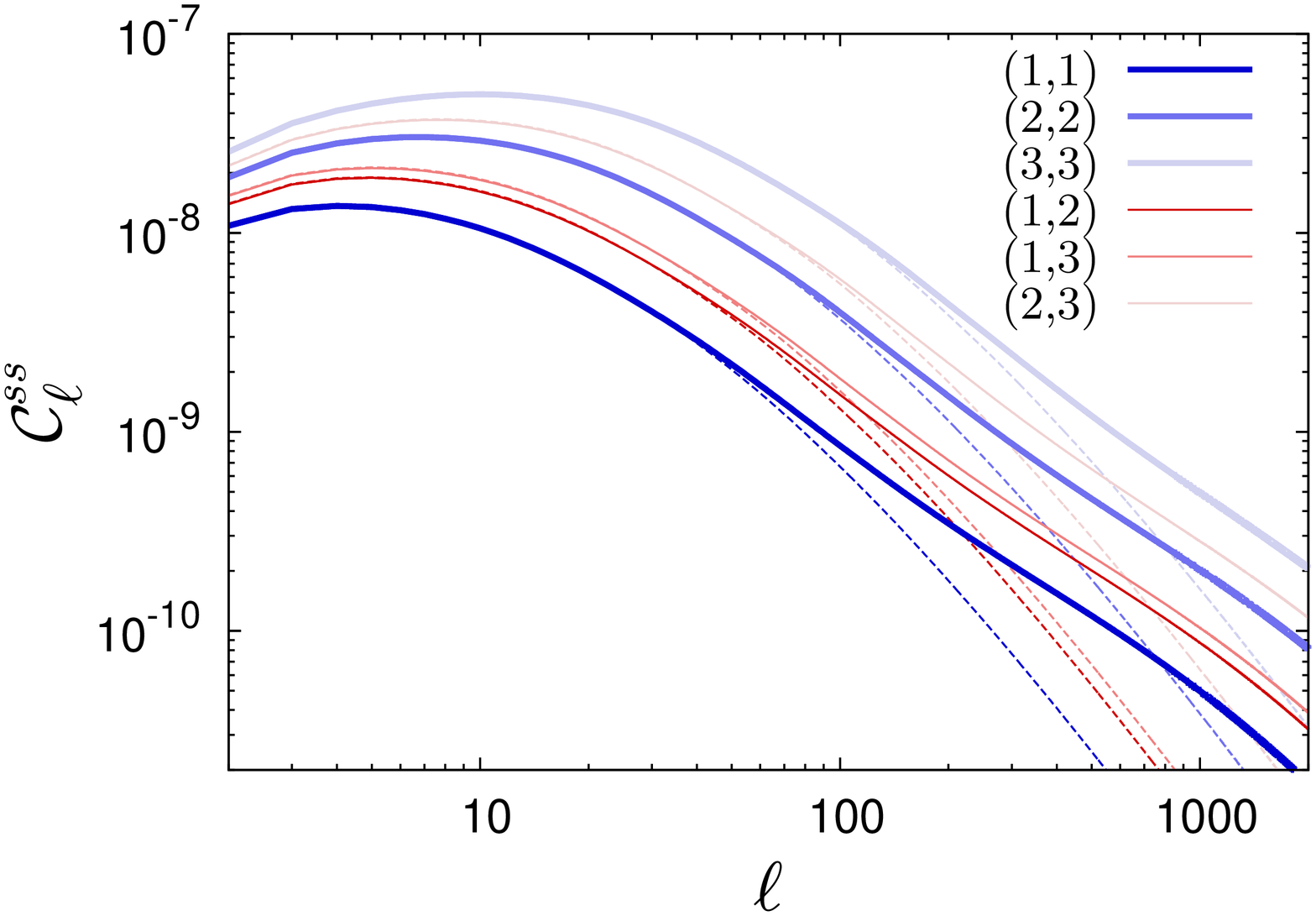}
\includegraphics[height=.66\textwidth,angle=270]{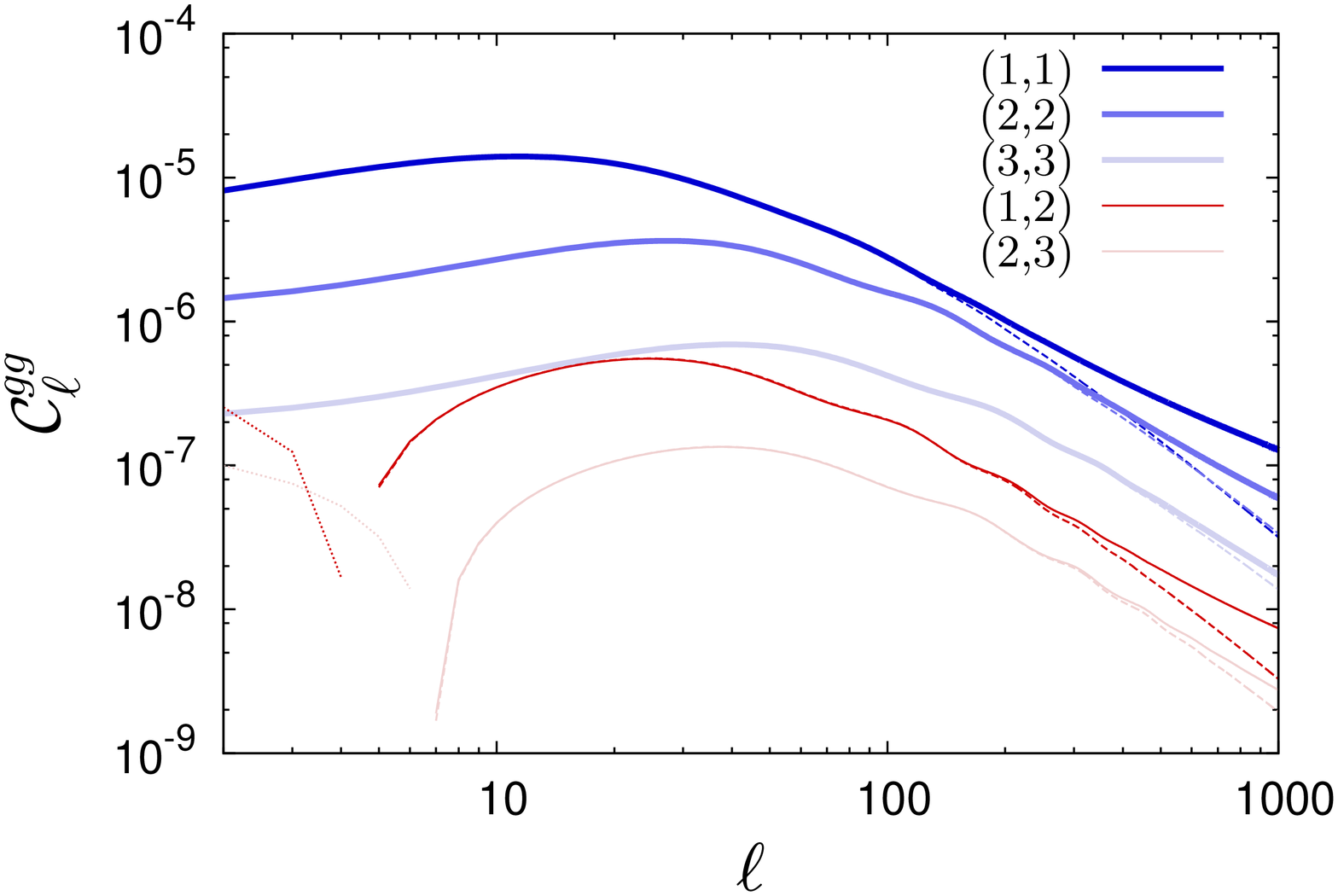}
\caption{Angular power spectra for three redshift bins (bin~1:~$z \in [0,0.61]$, bin~2:~$ z\in [0.61,1.03]$, bin~3:~$z \in [1.03,3]$).  Thick solid lines are the bin-autocorrelation spectra, thin solid lines bin-cross-correlations, both include \texttt{HaloFit} nonlinear corrections.  The corresponding linear theory spectra are plotted as dashed lines. {\it Top:} cosmic shear angular power spectra. {\it Bottom:}  galaxy angular power spectra.  At low multipoles $\ell$, the galaxy signals in different redshift bins are anticorrelated. The $(1,3)$-cross-spectrum is strongly suppressed because the corresponding bins are not adjoining, and is therefore not shown in the plot.\label{fig:spectra}}
\end{figure}

\begin{figure}[t]
\center
\includegraphics[height=.66\textwidth,angle=270]{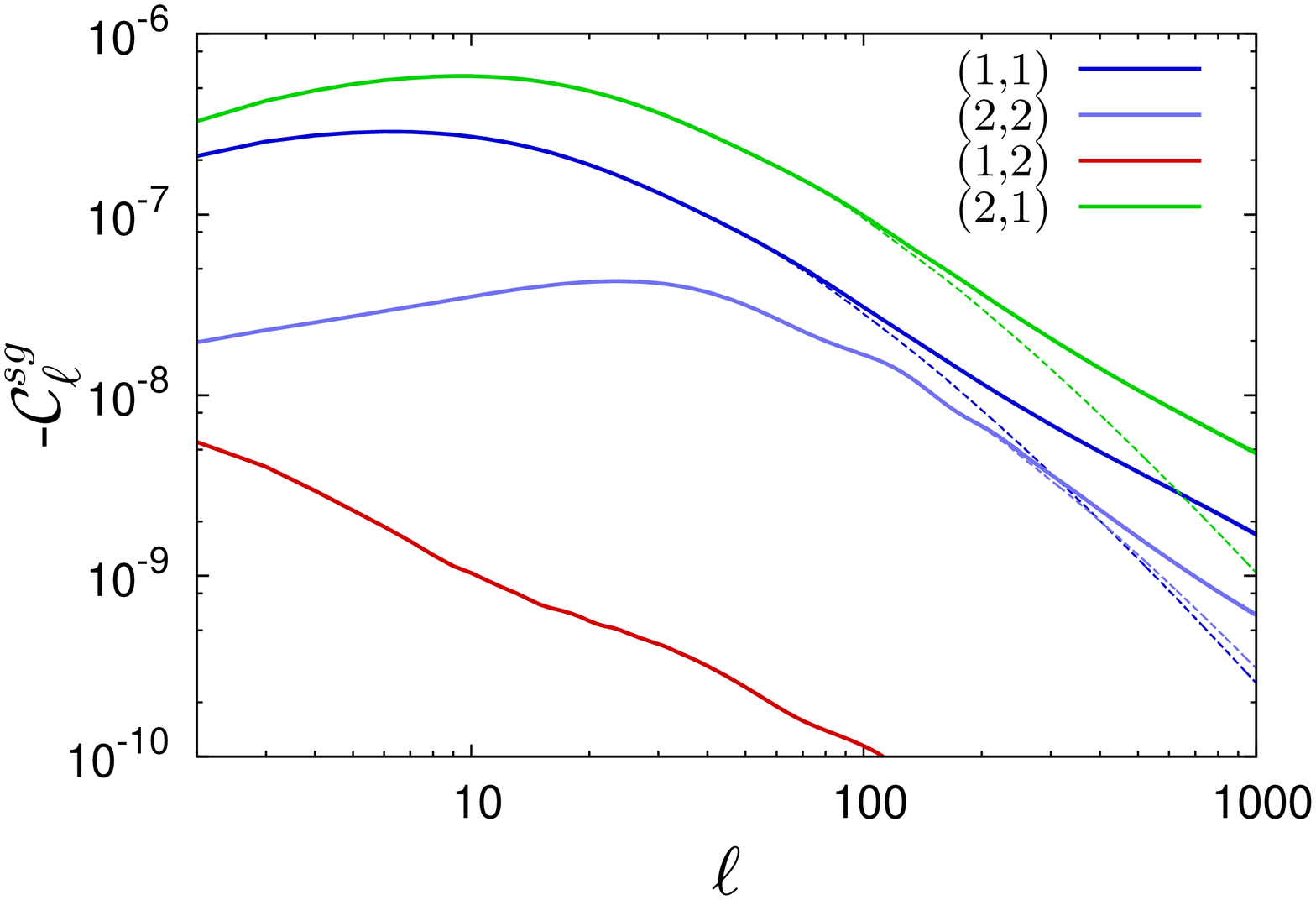}
\caption{Angular shear-galaxy cross-power spectra for two identical redshift bins in shear and galaxies (bin~1:~$z \in [0,0.61]$, bin~2:~$ z\in [0.61,1.03]$).  The first number denotes the index of the shear redshift bin, the second the index of the galaxy bin.  Note that unlike the ss and gg bin-cross-correlations, the sg cross-correlations are not symmetric in the redshift bin indices.  Solid lines include \texttt{HaloFit} nonlinear corrections, while dashed lines are the corresponding linear theory spectra. \label{fig:xspectra}}
\end{figure}

Note that, at first glance, our definitions of the galaxy and shear power spectra in equations~(\ref{eq:cl}) to~(\ref{eq:source2}) appear to differ from their conventional forms found in the literature (see, e.g.,~\cite{Zhan:2008jh}).  This difference can be traced to the following:
\begin{enumerate}
\item  We have not implemented {\it a priori} the Limber approximation, which consists in simplifying the source function integrals~(\ref{eq:source1}) and~(\ref{eq:source2}) to
\begin{align}
\label{eq:source1limber}
  \mathcal{S}_{\ell,i}^{\mathrm s} & \to  -2 \ \sqrt{\frac{\ell (\ell^2 - 1) (\ell+2)}{4}}  \left.  \sqrt{\frac{\pi}{2 \ell k^2} } \,  \mathcal{W}_i^{\rm s}(\chi)  \, T_\Psi(k,\eta_0 - \chi)\right|_{\chi=\ell/k}, \\
 \label{eq:source2limber}
  \mathcal{S}_{\ell,i}^{\mathrm g} & \to \left.  \sqrt{\frac{\pi}{2 \ell k^2} }\, \mathcal{W}_i^{\rm g} (\chi) \, T_\delta(k,\eta_0 - \chi)  \right|_{\chi = \ell/k}.
\end{align}
We do however use the  approximation in our numerical computations at \mbox{$\ell > 1000$}.

\item We have opted to highlight the fact that shear probes the {\it metric perturbations}, while galaxy distribution probes the {\it matter density fluctuations}.  While these quantities are correspondent under the Einstein equation in  simple $\Lambda$CDM-type models,
the distinction may be important in modified gravity scenarios, as well as in models that predict dark energy clustering.

\item The factor  $ \sqrt{\ell (\ell^2 - 1) (\ell+2)/4}$ in the shear source function arises from the fact that we are computing the $\epsilon$-mode shear without resorting to
the flat-sky approximation~\cite{Hu:2000ee}.

\end{enumerate}
For illustration,  figures~\ref{fig:spectra} and~\ref{fig:xspectra} show several sample spectra for our benchmark cosmological model (see section~\ref{sec:fiducial} for the cosmological parameter values).

\subsection{Source galaxy redshift distribution\label{sec:source}}

We model the redshift distribution of the source galaxies~(\ref{eq:galredshift}) using
\begin{equation}
{\rm d}n(z)/{\rm d}z \propto z^2 \exp(-(z/z_0)^\beta),
\label{eq:galdist}
\end{equation}
where  $z_0$ and $\beta$ are parameters specific to the survey.  This is a universal form for any magnitude-limited survey. The  factor $z^2$, together with the infinitesimal redshift element~$\di z$,  associates a volume to a given solid angle, and the exponential cut-off truncates the distribution  beyond the distance at which a typical survey galaxy has an apparent magnitude equivalent to the limiting magnitude of the survey~\cite{Kaiser:1996tp,Koo:1996ix,Hu:1999ek}.   In our analysis we set $\beta = 1$ and choose $z_0 = 0.3$ so that we recover {\sc Euclid}'s projected median survey redshift of roughly~0.8.  Note however that our final results (i.e., the estimated errors on various cosmological parameters) do not depend strongly on the details of the source galaxy redshift  model, as long as it follows qualitatively the shape described by the  expression~(\ref{eq:galdist}) and has the specified median redshift.

\section{Measurement errors\label{sec:errors}}

Measurements of the galaxy and the shear angular power spectra $C^{XY}_{\ell, ij}$ are subject first and foremost to cosmic variance, which poses an irreducible uncertainty
of $\Delta C_\ell/C_\ell = \sqrt{2/(2 \ell +1)}$ on how well the power spectra can be determined even in a perfectly controlled observation.  Cosmic variance is folded into our analysis at the level of the likelihood function, to be discussed in section~\ref{sec:likelihood}.  Other sources of measurement errors need to be dealt with on a more fundamental level.  We discuss these below.

\subsection{Shot noise\label{sec:shotnoise}}

A finite surface density of source galaxies limits the largest multipoles at which $C^{XY}_{\ell,ij}$ can be accurately measured. This finite surface density appears as a shot-noise term
\begin{equation}
\label{eq:shotnoise}
\delta C_{{\rm noise},ij}^{XY} = \delta_{ij} \delta_{XY} (\Xi^{X}_{i})^2 n_i^{-2},
\end{equation}
which contributes to the observed angular power spectra in addition to the ``signal''~$C^{XY}_{\ell,ij}$.  Here,
$\Xi^{\rm g}_i=1$ for the galaxy power spectrum measurement,  $\Xi^{\rm s}_i = \langle \gamma^2 \rangle^{1/2}$ corresponds to the root-mean-square ellipticity of the source galaxies for cosmic shear ($\langle \gamma^2 \rangle ^{1/2} \sim 0.16$ for {\sc Euclid}~\cite{Laureijs:2011mu}), and
the Kronecker deltas $\delta_{ij}$ and $\delta_{XY}$ ensure that the noise term contributes only to  autocorrelation spectra.
The source galaxy surface densities $n_i$ in the redshift bin $i$ can be computed as per equation~(\ref{eq:ni}), under the condition that $\sum_i n_i$ corresponds to the total galaxy surface density for the survey in question: for {\sc Euclid}, this is  anticipated to be 40~galaxies/arcmin$^2$~\cite{Laureijs:2011mu}.

\subsection{Photometric redshift uncertainties}

The {\sc Euclid} photometric survey does not have direct redshift information other than through filters. However, a combination of the spectroscopic survey and ground-based near-infrared photometry will be used to calibrate the photometric redshift measurements, and a precision of $\sigma(z) \sim 0.03 (1+z)$ has been shown to be achievable \cite{Laureijs:2011mu}.  Because {\sc Euclid} will not be able to measure in the near-infrared,
the addition of ground-based observations will also be crucial for minimising systematic biasing of the photometric redshift estimates~\cite{Abdalla:2007uc}.

We model the photometric redshift  uncertainty as a simple Gaussian error of standard deviation $\sigma(z)$ and with no bias.  This error can be folded into the window functions $\mathcal{W}^X_i$ as described in reference~\cite{Ma:2005rc}.
While $\sigma(z)$ of the magnitude expected for {\sc Euclid} is not critical for parameter estimation in $\Lambda$CDM-type models,
it can lead to a significant degradation in the survey's sensitivity to, for example, a time-dependent equation of state parameter in dynamical dark energy models~\cite{Ma:2005rc}.

In principle, redshift space distortions due to the collapse of large structures further affect the window functions.  However, since these are of order $10^3~{\rm km} \ {\rm s}^{-1}$, corresponding to $\delta z \sim 0.003$, they will be subdominant to the effect of the photometric redshift error at all but the smallest redshifts.  We therefore ignore them in the following.

\subsection{Shear-specific systematic errors}

\paragraph{Shear measurement errors}

An additional source of error comes from the experimental extraction of shear from the galaxy images. For example, surveys like {\sc Euclid} must be calibrated against large-scale simulation templates, and errors in the calibration result in an overall multiplicative error in the shear estimate.
Spurious contribution to the shear arises also from, e.g., an asymmetric point spread function, resulting in a so-called additive error.
A large and concerted effort is currently ongoing to quantify the magnitude of multiplicative and additive systematics, and to devise ways to correct for them (see, e.g.,~\cite{Heymans:2005rv,Massey:2006ha,Bridle:2008iv,Refregier:2012kg}). While it is currently not obvious that the bias introduced by multiplicative and additive errors is correctable to the level where it is no longer a limiting factor for the {\sc Euclid} survey~\cite{Refregier:2012kg}, we will assume that the correction is possible, and ignore these systematics in our analysis.

\paragraph{Intrinsic alignment}

Intrinsic alignment refers to the spurious shear correlation signal that arises (i) when the source galaxies are close in physical space and their intrinsic ellipticities (i.e., their orientations) becomes aligned by the tidal forces of  the gravitational field, and (ii) when the tidal-force field experienced by one source galaxy also causes the lensing of the images of higher-redshift galaxies at close projected distances, thereby leading to an alignment between galaxy images at different redshifts.
See, e.g.,~\cite{Croft:2000gz,Heavens:2000ad,Hirata:2004gc,Hirata:2007np} for detailed discussions of intrinsic alignment.

Because intrinsic alignment affects those galaxies that are close in either physical or projected distance, for a fixed multipole $\ell$ we expect its contribution to the signal to increase with decreasing redshifts.   For this reason, one may contemplate the exclusion of redshift bins below a certain threshold from the analysis.
However, because of the strong correlation with neighbouring bins, the very lowest redshift bins for shear are in most cases not crucial for parameter estimation, except perhaps in models that evolve rapidly at very late times (as might be the case in models with a time-dependent dark energy equation of state).
The removal of intrinsic alignment effects in the {\sc Euclid} data also appears to be possible (see, e.g.,~\cite{Blazek:2012hm,Kirk:2011aw}).   We therefore do not cut
the analysis at low redshifts, but rather keep the minimum $z$ at zero.

\section{Mock data generation\label{sec:mockdatageneration}}

Generalising the approach of reference~\cite{Perotto:2006rj}, we construct mock galaxy and shear angular auto- and cross-correlation spectra based on the specifications and error expectations of the {\sc Euclid} survey.  We describe the details of the construction below.  The synthetic data sets to be used in our parameter error forecast analysis are summarised in section~\ref{sec:datasets}.

\subsection{Mock angular power spectra}

For a full-sky observation, a fully random mock data synthesis requires that we pick a fiducial cosmological model, and compute for it the total theoretical covariance matrix
\begin{equation}
\label{eq:totalcl}
\bar{C}^{XY}_{\ell, ij} = C^{XY}_{\ell, ij}+ \delta C^{XY}_{{\rm noise},ij},
\end{equation}
where $C^{XY}_{\ell, ij}$ is now understood to include all error contributions discussed in section~\ref{sec:errors} except shot noise, which is encoded separately in
$\delta C^{XY}_{{\rm noise},ij}$.
A random realisation of the observables $a_{\ell m}^{X,i}$ obeying $\bar{C}^{XY}_{\ell, ij}$
can then be obtained following the multivariate random number generation procedure described
in~\cite{Perotto:2006rj,Hamann:2007sk}, such that
$\hat{C}^{XY}_{\ell, ij} \equiv (2 \ell +1)^{-1} \sum_{m=-\ell}^\ell a_{\ell m}^{X,i *} a_{\ell m}^{Y,j}$ gives the mock angular power spectra.  The procedure is repeated multiple times  to generate an ensemble of realisations.

However, for our purpose of estimating {\sc Euclid}'s sensitivity to various cosmological parameters (i.e., finding the size of the error bars), it is not necessary to generate a full ensemble of mock data; it suffices to set the mock angular power spectra $\hat{C}^{XY}_{\ell, ij}$ to be equal to the total covariance matrix $\bar{C}^{XY}_{\ell, ij}$ of the fiducial model, and hence perform the parameter estimation directly with the fiducial angular power spectra themselves.

\subsection{Partial sky coverage}

In principle, the above mock data generation procedure applies only in the case of a full-sky observation.  The {\sc Euclid} photometric survey, however, will cover only some 15,000 square degrees of the sky, equivalent to  an effective sky fraction of  $f_{\rm sky} = 0.3636$.
Any realistic survey that does not have full sky coverage necessarily yields angular power spectra $C^{XY}_{\ell, ij}$ that are correlated between $\ell$-values because
of any incomplete $(\theta,\phi)$-basis on which to perform the spherical harmonic decomposition.

The precise correlations between the multipoles will depend on the survey masks to be used.  However, for surveys such as the {\sc Euclid} photometric survey which have a reasonably symmetric mask (mainly due to a cut at low galactic latitudes), the loss of information due to the correlations between multipoles can be modelled approximately by augmenting the cosmic variance uncertainty as $\Delta C_\ell \to \sqrt{1/f_{\rm sky}} \  \Delta C_\ell$.  This does not affect the mock data generation procedure, but will have to be implemented in the parameter inference analysis at the level of the likelihood function (see section~\ref{sec:likelihood}).

\subsection{Redshift binning}
\label{sec:zbinning}

Although our source galaxy redshift model~(\ref{eq:galdist}) contains a built-in exponential cut-off at high redshifts, in practice we limit our analysis to the redshift interval $z \in [0,3]$ because {\sc Euclid} will observe very few galaxies beyond $z \sim 3$. This happens both because of the magnitude limitations of the survey and because the peak emission wavelengths are shifted out of the detectable region \cite{Laureijs:2011mu}.

A simple redshift binning scheme would be to slice the complete redshift range into bins of equal widths~$\Delta z$.  However, because ${\rm d} n(z)/{\rm d}z$ is not a constant in $z$, such a binning scheme would lead to unequal galaxy surface densities $n_i$ and hence disparate noise properties (see section~\ref{sec:shotnoise}) in different redshift bins.  Here, we adopt a different binning scheme, and choose our redshift slices in such a way that all bins contain the same~$n_i$.  Compared with a constant-$\Delta z$ binning, this scheme leads to narrower bins around the peak of ${\rm d} n(z)/{\rm d}z$, and wider ones in the tails.   By requiring the
same noise contributions in all bins, we have thus avoided having ``useless'' bins that are swamped by noise.
Additionally, this method is not very sensitive to the choice of the high-redshift cut-off, as long as it is far in the tail of the source galaxy redshift distribution.

\begin{figure}[t]
\center
\includegraphics[height=.68\textwidth,angle=270]{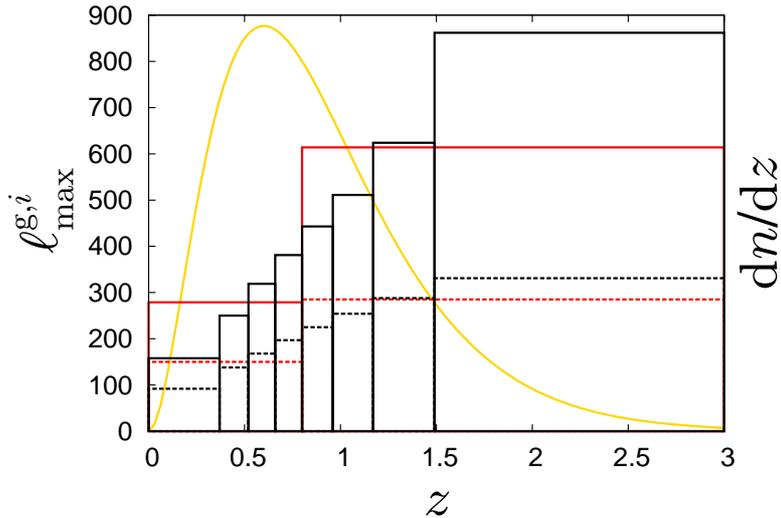}
\caption{Two examples illustrating our redshift binning scheme and the corresponding redshift-dependent $\ell_{\rm max}^{\rm g}$ for mock galaxy data.  Black lines denote eight redshift bins, while red lines represent a two-bin scheme.  Solid lines correspond to $\epsilon_{\rm nl} = 1$, and dashed lines represent the linear-only case with $\epsilon_{\rm nl} = 0.1$.  Also plotted in gold is the differential source galaxy surface density from equation~(\ref{eq:galdist}), with an arbitrary normalisation.\label{fig:lmax}}
\end{figure}

Figure~\ref{fig:lmax} shows two examples of our equal $n_i$ redshift binning  scheme, one for eight bins and one for two bins.

\subsection{Nonlinearity, scale-dependent bias, and the maximum angular harmonic\label{sec:nonlinear}}

The contribution of nonlinearities on small scales is an important
issue for cosmological parameter inference from large-scale structure
observations. At the most basic level, the breakdown of linear
perturbation theory as a description of the gravitational clustering
of (dark) matter at large wavenumbers has observational consequences
for all large-scale structure surveys.  This kind of nonlinearity is
generally calculable by way of $N$-body simulations, at least on the
mildly nonlinear scales.  At a more advanced level, observational
probes that make use of tracers such as galaxies suffer in addition
the presence of a scale-dependent bias, which is a considerably more
severe form of nonlinearities because it cannot be predicted from
first principles.  Measures need to be taken to ensure that we account
for these nonlinear effects properly in our analysis.

\paragraph{Shear power spectrum}

Cosmic shear probes directly the metric perturbations, so that tracer bias is not a concern.  The main source of nonlinearities on the observable scales is nonlinear gravitational effects, although dissipative baryon processes could conceivably contribute a few percent to the power spectrum at $\ell> 2000 \to 3000$.
 At very high multipoles non-Gaussian effects from nonlinearities also become important, and an analysis in terms of the power spectrum breaks down. However, based on ray tracing, reference~\cite{Hilbert:2008kb} finds this not to be a dominant effect unless the analysis is taken to beyond $\ell \sim 10^4$.

In our analysis, we  impose a redshift-independent cut-off at $\ell_{\rm max}^{\rm s}=2000$.  The specific value is unimportant, because for a {\sc Euclid}-like survey,
shot noise always dominates over the signal  before this cut-off is reached.  Although this relatively low $\ell_{\rm max}^{\rm s}$ precludes any strongly nonlinear data
from entering into our analysis, it is nonetheless important that we correct for nonlinearities on the mildly nonlinear scales.
In practice this means that when constructing our mock shear data, we compute our fiducial shear power spectrum assuming the nonlinear corrections of~\texttt{HaloFit}~\cite{Smith:2002dz}, including the recent recalibration of reference~\cite{Bird:2011rb}.  We take the view that these nonlinear corrections are calculable to the accuracy required to maintain an unbiased estimation of the cosmological parameters
 (or will be by the time real {\sc Euclid} measurements become available), and analyse the mock shear data assuming  \texttt{HaloFit} provides the correct nonlinear modelling.
See, e.g.,~\cite{Hearin:2011bp}  for a recent study of the accuracy requirement on the nonlinear matter power spectrum in the context of dark energy parameter determination.

\paragraph{Galaxy power spectrum}

Parameter inference from, e.g., the three-dimensional power spectrum of the SDSS LRG galaxies usually require that we cut off the spectrum at a wavenumber of
$k \sim 0.1 \to 0.2 \, h \ {\rm Mpc}^{-1}$ in order to limit the effect of scale-dependent galaxy bias.  Alternatively, one may fold in the scale dependence in a bias model and marginalise over the model's nuisance parameters, as was done in, e.g., reference~\cite{Hamann:2008we} using the $P$-model, and reference~\cite{Cole:2005sx} using the $Q$-model.

Here, we  tackle the problem of scale-dependent bias by introducing in the  galaxy angular power spectrum
a redshift bin-dependent cut-off  $\ell_{\rm max}^{{\rm g},i}$.
Our procedure is motivated by two premises.  Firstly, the sensitivity of the measurement (which, at large scales, is limited by cosmic variance) determines the required accuracy of the theoretical prediction.  Secondly, the relative nonlinear correction to the matter power spectrum serves as an indicator of the scale at which
 nonlinear biasing effects start to become relevant.  We therefore define $\ell_{\rm max}^{{\rm g},i}+1$ as the smallest $\ell$ value that satisfies the inequality
\begin{equation}
\epsilon_{\rm nl}  \left( 1 + \sqrt{\frac{2}{2 \ell + 1}} \right) \ge \left. \frac{\mathcal{P}_{\rm nl}(k,\bar{z}_i)}{\mathcal{P}_{\rm lin}(k,\bar{z}_i)} \right|_{k = \ell/\chi(\bar{z}_i)},
\end{equation}
where $\epsilon_{\rm nl}$ is a nonlinear tolerance factor, $\bar{z}_i$ the mean redshift of bin~$i$, $\mathcal{P}_{\rm lin}$  the linear matter power spectrum, and $\mathcal{P}_{\rm nl}$ the nonlinear matter power spectrum corrected with~\texttt{HaloFit}.  We consider two choices of the tolerance factor~$\epsilon_{\rm nl}$:
\begin{itemize}
\item $\epsilon_{\rm nl} = 0.1$; this  ``linear only'' scenario  completely eschews any remotely nonlinear region.

\item $\epsilon_{\rm nl}=1$; this slightly more optimistic case requires a modest, but not unrealistic amount of nonlinear modelling, and is our default choice.

\end{itemize}
Generally, the higher the redshift, the smaller the nonlinear correction for a given length scale.  At the same time, however, going to higher redshifts also changes the projection;  the same scale is now projected to a higher~$\ell$, where cosmic variance is less so that the demand for accuracy of the power spectrum higher.
In practice, though, the wavenumber corresponding to $\ell_{\rm max}^{{\rm g},i}$ does not depend strongly on redshift; we find $\sim 0.14 \to 0.16$~Mpc$^{-1}$ in the default case, and $\sim 0.06 \to 0.09$~Mpc$^{-1}$ in the linear only scenario.

\paragraph{Shear-galaxy cross-correlation}

As discussed above, the scale-dependent bias for the galaxy auto-spectrum very quickly erases information at wavenumbers $k$ larger than a certain threshold.
Shear-galaxy cross-correlation is in principle slightly less sensitive to this effect, since  galaxy data contribute only to half of the measure.
However, the corresponding threshold in $k$-space turns out to be pushed back only by a small amount.  To see this,
take as an example the $P$-model introduced in~\cite{Hamann:2008we}, which describes the scale-dependent bias as a super-Poisson shot noise, i.e., a component with no $k$-dependence. On scales where the bias contribution is important, the power spectrum typically drops as $k^n$, with $n \lesssim -1.5$. Therefore, if  scale-dependent bias corrections become relevant for the galaxy auto-spectrum at $k_{\rm bias}$, then the corresponding threshold for the cross-spectrum should be at $2^{-1/n} k_{\rm bias} \simeq 1.5\ k_{\rm bias}$, which is a marginal shift.  For this reason, we take for simplicity the same $\ell_{\rm max}$ for the cross-correlation as for the galaxy auto-correlation.

\subsection{Synthetic {\sc Euclid} data\label{sec:datasets}}

We summarise here the mock data sets we generate for the {\sc Euclid} photometric survey.

\begin{itemize}
\item A weak lensing shear auto-spectrum, $C_{\ell,ij}^{\rm ss}$, where the multipole $\ell$ runs from 2 to $\ell_{\rm max}^{\rm s}=2000$ independently of redshift.
The indices $i,j \in [1,N_{\rm s}]$ label the redshift bin, where the redshift slicing follows the scheme described in section~\ref{sec:zbinning}.

\item  A galaxy auto-spectrum, $C_{\ell,ij}^{\rm gg}$, where the multipole runs from 2 to $\ell^{{\rm g},i}_{\rm max}$, and the indices $i,j \in [1,N_{\rm g}]$ label the redshift bin.
The maximum multipole $\ell^{{\rm g},i}_{\rm max}$ is redshift-dependent because the scale at which scale-dependent galaxy bias becomes important varies with redshift, as discussed in section~\ref{sec:nonlinear}.

\item A shear-galaxy cross-spectrum, $C_{\ell,ij}^{\rm sg}$, where $i \in [1,N_{\rm s}]$ labels the shear redshift bin and $j \in [1,N_{\rm g}]$ the galaxy redshift bin. For the cross-spectrum, the maximum $\ell$ values are always given by $\ell^{{\rm g},j}_{\rm max}$ of the galaxy redshift bin $j$.
\end{itemize}

\section{Forecasting\label{sec:forecasting}}

Having outlined the mock data generation procedure, we are now in a position to describe our parameter error forecast for the {\sc Euclid} photometric survey.
The forecast is based on the likelihood method, in which we construct a likelihood function of the mock data and then explore its implications using Bayesian inference techniques.

\subsection{Model parameter space\label{sec:fiducial}}

As a simple example, we consider a $\Lambda$CDM model extended with a nonzero neutrino mass specified by seven parameters:
\begin{equation}
\label{eq:model}
\Theta = (\omega_{\rm b},\omega_{\rm dm},h,A_{\rm s},n_{\rm s},z_{\rm re},f_\nu),
\end{equation}
where $\omega_b = \Omega_b h^2$ is the physical baryon density, $\omega_{\rm dm} = \Omega_{\rm dm} h^2$ the physical dark matter (cold dark matter and massive neutrinos) density, $h = H_0/(100 \, {\rm km} \, {\rm s}^{-1} \, {\rm Mpc}^{-1})$ the dimensionless Hubble parameter, $A_{\rm s}$  the amplitude of primordial scalar fluctuations, $n_{\rm s}$ the associated scalar spectral index, $z_{\rm re}$ the redshift of reionisation, and the neutrino density fraction is defined here as $f_\nu = \omega_\nu/\omega_{\rm dm}$. Our  fiducial model is defined by  the parameter values
\begin{equation}
\label{eq:fidvalues}
\Theta_{\rm fid} = (0.022, 0.1126228, 0.7, 2.1 \times 10^{-9}, 0.96, 11, 0.00553),
\end{equation}
which we use to generate the mock data sets described in section~\ref{sec:datasets}.

The fiducial neutrino fraction $f_\nu = 0.00553$ corresponds to a neutrino energy density $\omega_\nu = 0.0006228$, or
a sum of neutrino masses $\sum m_\nu = 94.1 \, \omega_\nu \, {\rm eV} \simeq 0.06 \, {\rm eV}$, roughly what one expects for a normal neutrino mass hierarchy in which the lightest mass eigenstate has a zero mass $m_1 \sim 0$.  We implement this fiducial neutrino energy density as being  contained entirely in one single neutrino species with mass $\sim 0.06$~eV, the other 2.046 species being exactly massless.
While this is formally inconsistent with the mass spectrum of the normal hierarchy, namely, $m_2 = m_1 + \sqrt{\Delta m_{12}^2} \simeq 0.009$~eV and $m_3 = m_1 + \sqrt{\Delta m_{12}^2} + \sqrt{\Delta m_{23}^2} \simeq 0.05$~eV,%
\footnote{We use $\Delta m_{12}^2 = 7.5 \times 10^{-5} \, {\rm eV}^2$ and $|\Delta m_{23}^2| = 2.3 \times 10^{-3} \, {\rm eV}^2$~\cite{Fogli:2012ua,Tortola:2012te}.
}
our simplified model is observationally indistinguishable from the correct one unless the survey sensitivity is much better than the mass splittings (see section~\ref{sec:hierarchy} and also reference~\cite{Lesgourgues:2004ps}).  Except in section~\ref{sec:hierarchy}, the following parameter estimation analysis will also assume only one massive neutrino and 2.046 massless neutrinos.

\subsection{Likelihood function\label{sec:likelihood}}

Because we are concerned only with parameter error estimation, not model comparison, we may define our likelihood function $\mathcal{L}$ in such a way that $\chi^2_{\rm eff} \equiv -2 \ln \mathcal{L}$ is exactly equal to zero for the fiducial parameter values.  If the covariance matrix of the data set has dimension~\mbox{$N \times N$}, this amounts to defining  $\chi^2_{\rm eff}$  as
\begin{equation}
\chi^2_{\rm eff} = \sum_{\ell} (2\ell+1) f_{\rm sky} \left[{\rm Tr}(\bar{\mathbf C}_{\ell}^{-1} \hat{\mathbf C}_{\ell}) + \ln \frac{{\rm Det}(\bar{\mathbf C}_{\ell})}{{\rm Det} (\hat{\mathbf C}_{\ell})} - N \right],
\end{equation}
where $\hat{\mathbf C}_\ell \equiv \hat{C}^{XY}_{\ell, ij}$ is the mock data covariance matrix, while $\bar{\mathbf C}_{\ell} \equiv \bar{C}_{\ell, ij}^{XY}$ is the total theoretical covariance matrix of equation~(\ref{eq:totalcl}).

From the three mock {\sc Euclid} data sets we generate, four different combinations are possible:
\begin{itemize}
\item Galaxy auto-spectrum only, $N=N_{\rm g}$.
\item Shear auto-spectrum only, $N=N_{\rm s}$.
\item Galaxy and shear auto-spectra, with no shear-galaxy cross-spectrum. In this case  $\chi^2_{\rm eff} = \chi^2_{\rm eff,g}+\chi^2_{\rm eff,s}$,
where the two contributions have  $N=N_{\rm g}$ and $N=N_{\rm s}$ respectively.
\item All galaxy and shear power spectra including shear-galaxy cross-correlation, $N=N_{\rm g}+N_{\rm s}$.
\end{itemize}

\subsection{Parameter inference}

We use the nested sampling algorithm~\cite{nestedsampling}, as implemented in the \texttt{MultiNest}~\cite{multinest1,multinest2} add-on to the \texttt{CosmoMC}~\cite{Lewis:2002ah} parameter estimation code, to infer constraints on the  parameters of our cosmological model~(\ref{eq:model}).

Because the full model parameter space is quite large and some of the parameters, especially $\omega_b$ and $z_{\rm re}$, are badly or completely unconstrained by large-scale structure observations, we use in addition a set of mock CMB $TT$, $EE$ and $TE$ data simulated following the procedure of~\cite{Perotto:2006rj}
according to the specifications of {\sc Planck}~\cite{Planck:2006aa}, using information from the 100, 143 and 217~GHz channels.  We do not model foreground signals, beam uncertainties, etc., but instead cut off the spectra at a maximum multipole of $\ell_{\rm max}^{\rm CMB} = 2000$ in both temperature and polarisation.

\subsection{Treatment of galaxy bias\label{sec:bias}}

While we have adopted a very conservative cut in the high multipoles of the galaxy angular power spectra in order to avoid the issue of scale-dependent galaxy bias, the
galaxy number density fluctuations still differ from the underlying matter density perturbations by a redshift-dependent linear bias factor.

A realistic linear bias model should be a continuous and monotonic function of redshift, one that is probably describable in terms of only two or three free parameters.  In the absence of a concrete model, however, we consider two extreme assumptions about the linear bias  when analysing our mock galaxy data:

\begin{enumerate}
\item In the most optimistic scenario, the linear bias factor is perfectly known, so that the galaxy power spectrum  measures directly the underlying matter power spectrum with no loss of information.   In this case, without loss of generality, we can set the linear bias factor in every redshift bin to $b_i=1$.

\item In the most pessimistic scenario, we have no knowledge of the linear bias at all, which makes it necessary to postulate $N_{\rm g}$ independent linear bias factors, one for each of the $N_{\rm g}$ redshift bins.   These $N_{\rm g}$ linear bias parameters are let to vary in our parameter inference, and marginalised at the end of the analysis.
Under this assumption, not only do we discard information coming from the normalisation of the galaxy density fluctuations, we also throw away information about the growth of structures as a function of redshift.

\end{enumerate}

\section{Results\label{sec:results}}

As discussed above,  given the survey specifications, there are a number of ways in which we may construct our mock data sets and subsequently analyse them.
For both cosmic shear and galaxies, we have the freedom to select the subdivision of the data according to redshift.  For galaxies,
we additionally need to specify a bin-dependent $\ell^{\rm g}_{\rm max}$ and how to deal with the galaxy bias.
The choices we make will eventually affect the parameter constraints derived from the data.

We begin this section by assessing the impact of our assumptions on cosmological parameter inference from the individual shear and galaxy mock data sets together with mock
{\sc Planck} CMB data.
After that, we consider all  data sets in combination, and determine the best parameter sensitivities attainable by {\sc Euclid}.

\subsection{Varying the number of redshift bins \label{sec:binning}}

Given our assumptions of an unbiased redshift determination and a
known photometric redshift uncertainty, one might expect that
increasing the number of redshift bins $N_{\rm bin}$ will always lead
to smaller uncertainties in parameter estimates, until the bin widths
become comparable to the photometric redshift error, or the galaxy
surface density too small to maintain a sizeable signal-to-noise
ratio.  In reality, however, the gain in information on cosmological
parameters from redshift binning often saturates well before this
point is reached.  This is understandable for parameters that have a
negligible effect on the growth of structures as a function of
redshift, e.g., the parameters that describe the primordial
perturbations.  However, because the growth function evolves slowly
and monotonically with redshift in most reasonable cosmologies,
signals in neighbouring redshift bins can become highly correlated, so
that even parameters that do have a strong impact on the growth
function may be subject to the said saturation effect.

\begin{figure}[t]
\center
\includegraphics[height=.48\textwidth,angle=270]{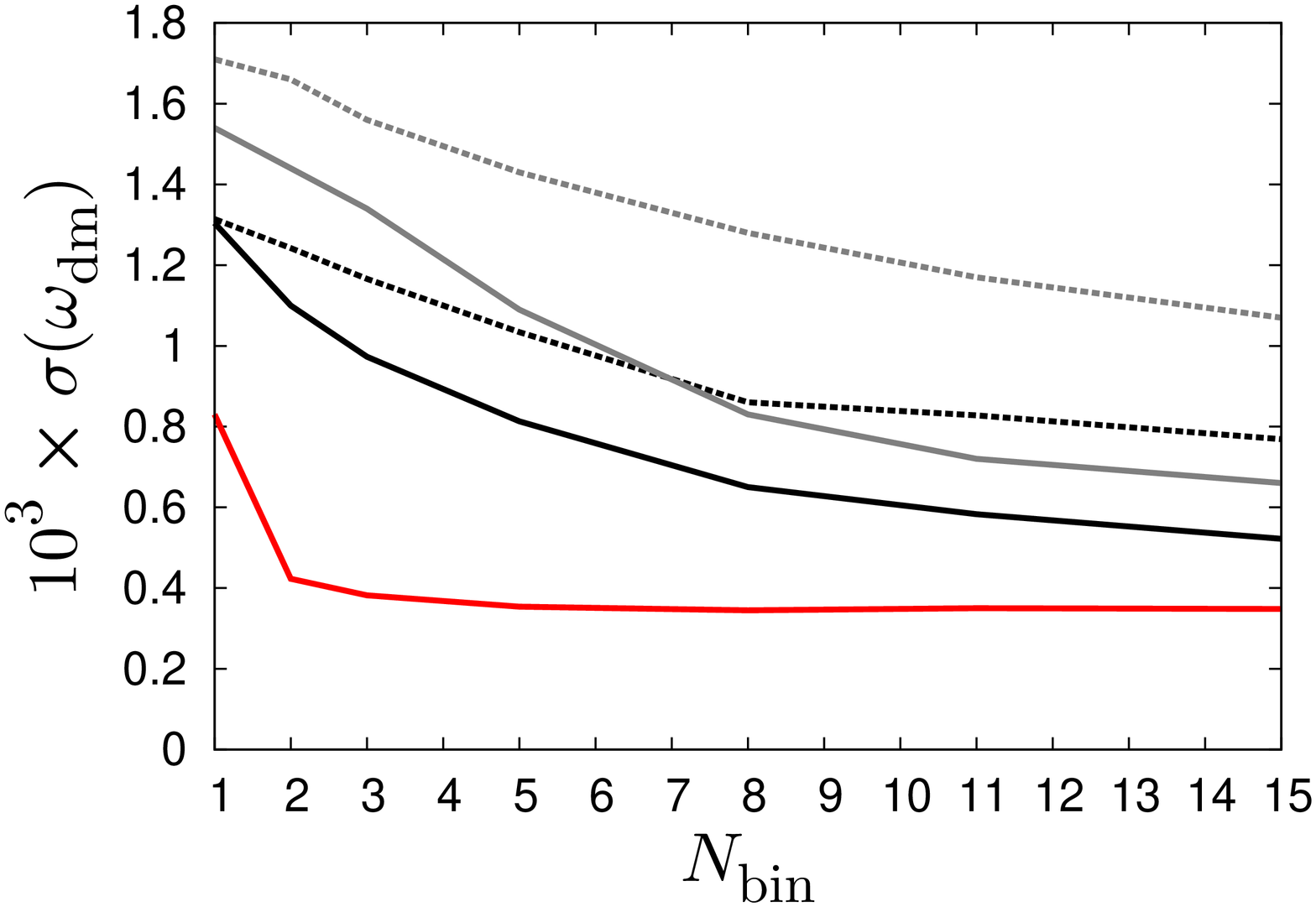}
\includegraphics[height=.48\textwidth,angle=270]{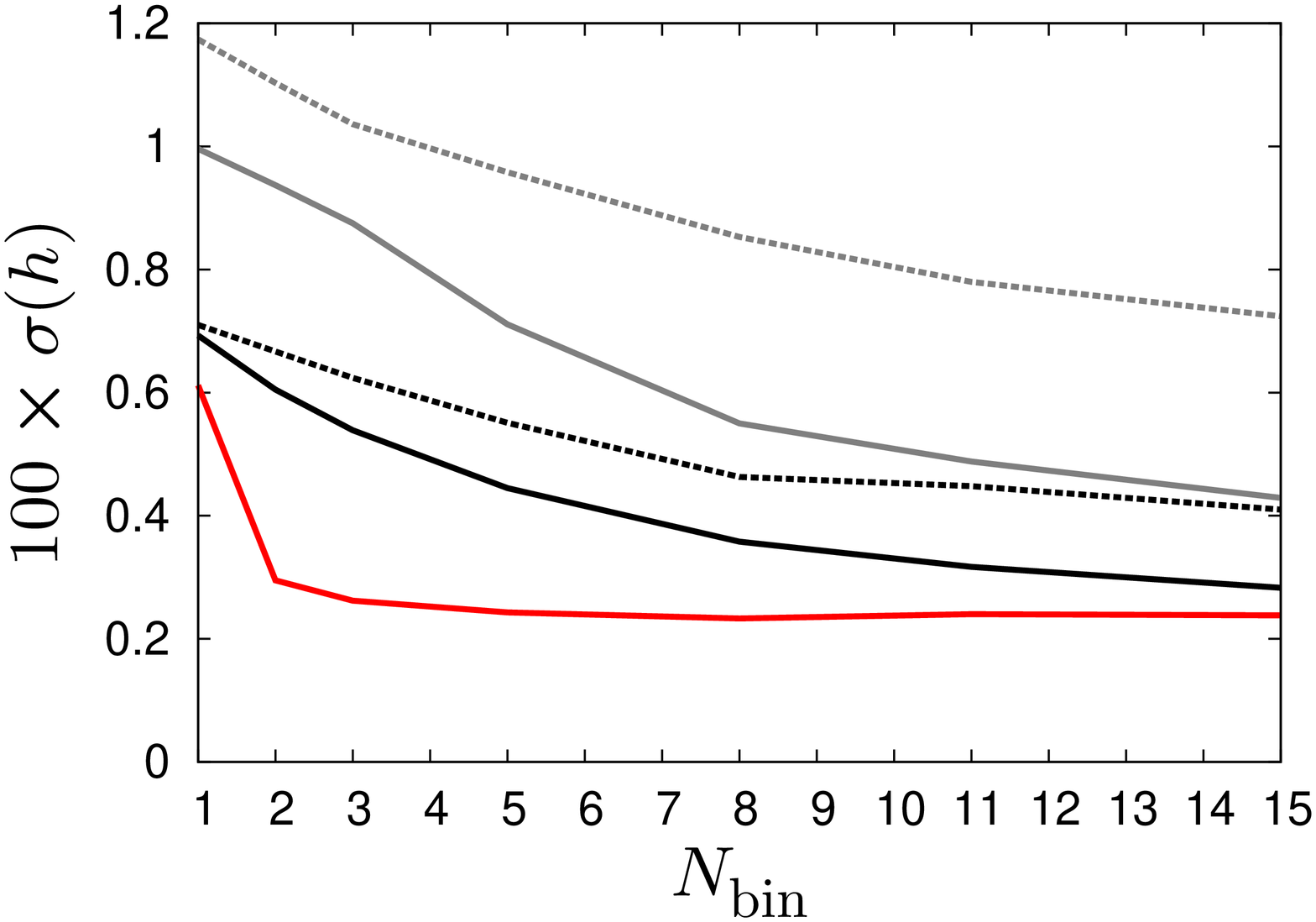}
\includegraphics[height=.48\textwidth,angle=270]{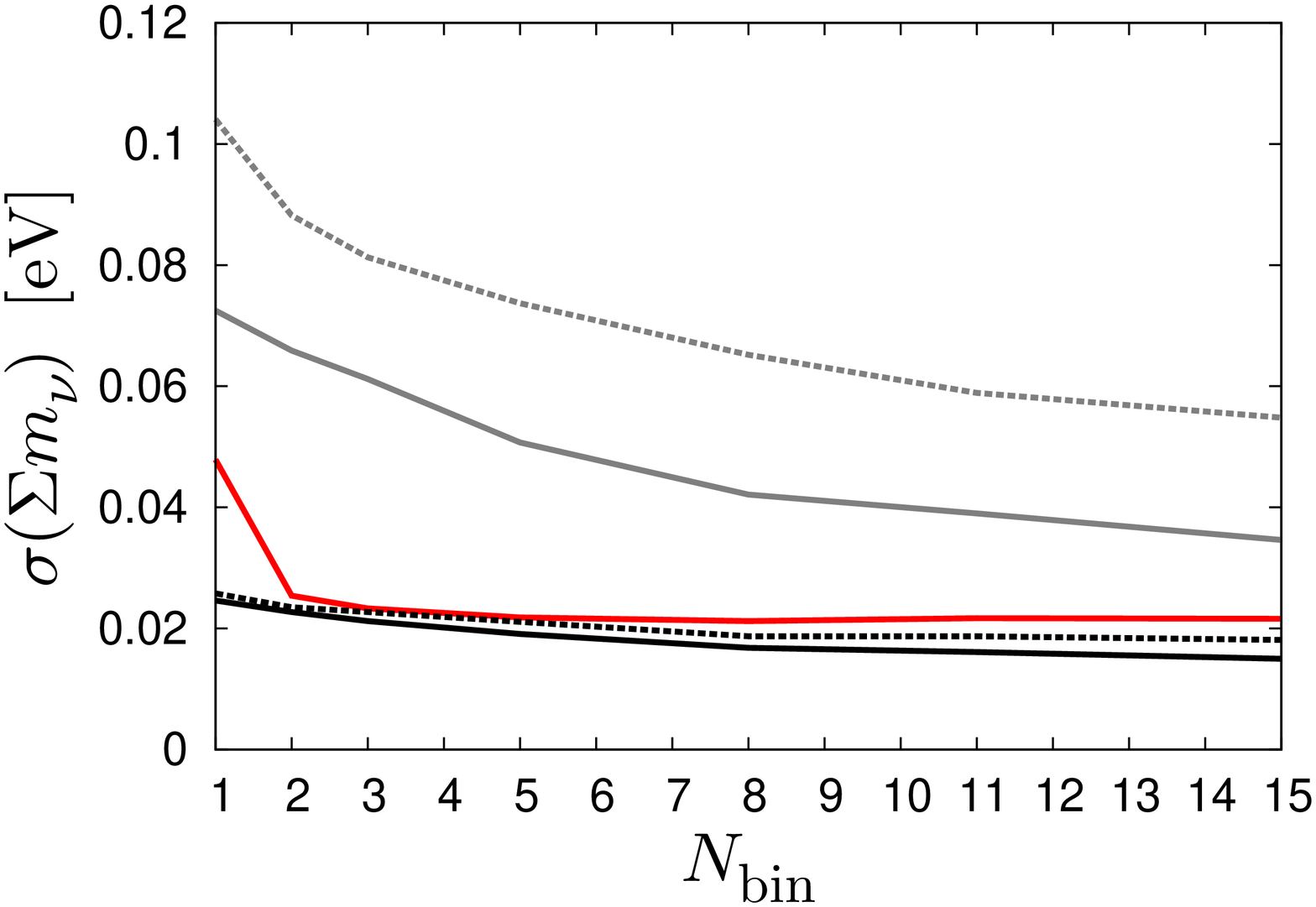}
\caption{Standard deviation of the marginalised posterior distribution for $\omega_{\mathrm dm}$ ({\it top left}), $h$ ({\it top right}) and  $\sum m_\nu$ ({\it bottom}), derived for various combinations of synthetic {\sc Planck} CMB data and  {\sc Euclid} data.  {\it Red:} {\sc Planck}+shear auto-spectrum. {\it Black/Grey:}  {\sc Planck}+galaxy auto-spectrum without/with marginalisation over $N_{\rm bin}$ bias parameters.  Dotted lines denote galaxy data on linear scales only ($\epsilon_{\rm nl}=0.1$), while solid lines include data on mildly nonlinear scales ($\epsilon_{\rm nl} = 1$).\label{fig:binning}}
\end{figure}

Figure~\ref{fig:binning} shows how the standard deviation of the
marginalised posterior probability distribution for $\omega_{\rm dm}$,
$h$, and $\sum m_\nu$ evolves as a function of the number of bins,
using {\sc Planck} CMB data combined with {\it either} shear {\it
  or} galaxy measurements, sliced in $N_{\rm bin}$ redshift bins.  The remaining four parameters
 of our seven-parameter model are almost completely constrained by {\sc Planck};
 adding shear or galaxy data to the inference leads only to a small reduction (up to $\sim 20\%$) in the error bars.

For {\sc Planck}+shear, saturation occurs already at two bins.  The improvement gained from going from one bin to two bins can be attributed to the fact that the latter improves the constraints on $\omega_{\rm dm}$ and $h$ through a measurement of the growth function, even if in a convoluted form, at two different redshifts.  (In $\Lambda$CDM-type cosmologies, the growth function is governed only by $\omega_{\rm m} \equiv \omega_{\rm dm} + \omega_{\rm b}$ and $h$; see equation~(\ref{eq:growth}).)
But a finer redshift binning adds little value to parameter inference in our simple seven-parameter model, because more bins would merely lead to an overdetermination of the two parameters that control the growth function.  For more complex model spaces, however, especially those with dynamical dark energy described by multiple parameters, employing three or more tomography bins may lead to yet more improvement in the parameter constraints (e.g.,~\cite{Ma:2005rc}).
Note that our forecast sensitivities to $\sum m_\nu$ for {\sc Planck}+shear
are somewhat better than the findings of~\cite{Hannestad:2006as}.  This is due to
a larger model parameter space and lower assumptions about the
survey performance adopted in the earlier analysis.

For {\sc Planck} combined with the galaxy auto-spectrum, there continues to be some improvement
up to around ten bins.    This is because a finer redshift binning in conjunction with our prescription for constructing the bin-dependent $\ell_{\rm max}^{\rm g}$ allows
us to measure the power spectrum up to higher multipoles (where cosmic variance is smaller) at high redshifts (see figure~\ref{fig:lmax}).
In turn, a higher $\ell_{\rm max}^{\rm g}$  lends a longer lever arm to probe the shape of the matter power spectrum and its small-scale amplitude.

\subsection{Bias modelling and small-scale cut-off}

As expected, bias marginalisation greatly degrades the galaxy auto-spectrum's sensitivity to cosmological parameters.  In the case of $\sum m_\nu$, for example, figure~\ref{fig:binning} shows a factor of two to three increase in the uncertainty.
This suggests that for the low fiducial neutrino masses considered in this work,
the full galaxy power spectrum, including normalisation information, constrains $\sum m_\nu$ predominantly through
a measurement of the spectrum's {\it absolute} amplitude on small scales,
rather than through a detection of the characteristic step-like feature at the neutrino free-streaming scale,
or  the impact of $\sum m_\nu$ on the growth function.

Further evidence  to support this point can be seen in figure~\ref{fig:binning}, when we
discard the mildly nonlinear scales in the galaxy angular power spectrum in our analysis,
 i.e., we decrease $\epsilon_{\rm nl}$ from
1 to 0.1.  If  the galaxy bias is exactly known, the degradation in the parameter uncertainties is
about 30 to 40\% for $\omega_{\rm dm}$ and $h$, and always less than 20\% for $\sum m_\nu$, with the loss of sensitivity being more
pronounced for finer redshift binnings.  However, if  the linear galaxy bias is unknown, then the additional loss of
information on the mildly nonlinear scales has a more dramatic effect on the parameter constraints.  For $\sum m_\nu$,
we find that between no nonlinear information and with nonlinear information, the reduction in sensitivity is about 40 to 60\%.  This tells us that
the extended ``lever arm'' afforded by the mildly nonlinear scales is more important for the power spectrum shape measurement than for the determination of the
small-scale spectrum amplitude.

\begin{table*}[t]
  \caption{Posterior standard deviations for the parameters $\omega_{\rm dm}$, $h$, and $\sum m_\nu$, derived from various combinations of data sets.  Here, ``c'' denotes {\sc Planck} CMB data, ``g'' galaxy auto-spectrum (11 bins), ``s'' shear auto-spectrum (2 bins), and  ``x'' shear-galaxy cross-correlation.   The subscripts ``l'' and ``b'' refer respectively to
a galaxy data set with $\epsilon_{\rm nl} = 0.1$, and full bias marginalisation.\label{tab:errors}}
\begin{center}
{\footnotesize
  \hspace*{0.0cm}\begin{tabular}
  {lccc} \hline \hline
  Data & $10^3 \times \sigma(\omega_{\rm dm})$ & $100 \times \sigma(h)$ & $\sigma(\sum m_\nu)/$eV \\       \hline
  c & 2.02 & 1.427 & 0.143 \\
  cs & 0.423 & 0.295 & 0.025 \\
  cg & 0.583 & 0.317 & 0.016 \\
  cg$_{\rm l}$ & 0.828 & 0.448 & 0.019 \\
  cg$_{\rm b}$ & 0.723 & 0.488 & 0.039 \\
  cg$_{\rm bl}$ & 1.165 & 0.780 & 0.059 \\
  csg & 0.201 & 0.083  & 0.011  \\
  csgx & 0.181 & 0.071 & 0.011 \\
  csg$_{\rm b}$ & 0.385 & 0.268 & 0.023 \\
  csg$_{\rm b}$x & 0.354 & 0.244 & 0.022 \\ \hline \hline
  \end{tabular}
  }
  \end{center}
\end{table*}

\subsection{Combining data sets}

Having studied the parameter sensitivities of CMB combined with {\it either} shear {\it
  or} galaxy data, we now proceed to consider shear and galaxy measurements in combination.
Based on the conclusions of section~\ref{sec:binning}, we choose to perform
the combined analysis using two shear and eleven galaxy redshift bins.
The results of this analysis are summarised in table~\ref{tab:errors}.

 Somewhat surprisingly, we find that as long as the galaxy bias is known,
constraints on $\omega_{\rm dm}$, $h$, and $\sum m_\nu$ in the combined analysis are
dramatically better than what one would na\"{\i}vely expect from looking at the sensitivities derived from the {\sc Planck}+shear (``cs'') and {\sc Planck}+galaxies (``cg'') analyses.
Most notably, a determination of $\sum m_\nu$ with an error as small as $\sigma(\sum m_\nu) = 0.011$~eV may be possible combining {\sc Planck}, shear and galaxy data (``csg'').
Such a  sensitivity would imply a detection of a nonzero neutrino mass at greater than five standard deviations, even
in the most pessimistic case of a massless lightest state in the normal mass hierarchy.  Galaxy bias marginalisation reduces the
sensitivity to  $\sigma(\sum m_\nu) = 0.023$~eV (``csg$_{\rm  b}$''), which still represents a $\sim 2.5 \sigma$ detection in the worst case.

\begin{figure}[t]
\center
\includegraphics[height=.48\textwidth,angle=270]{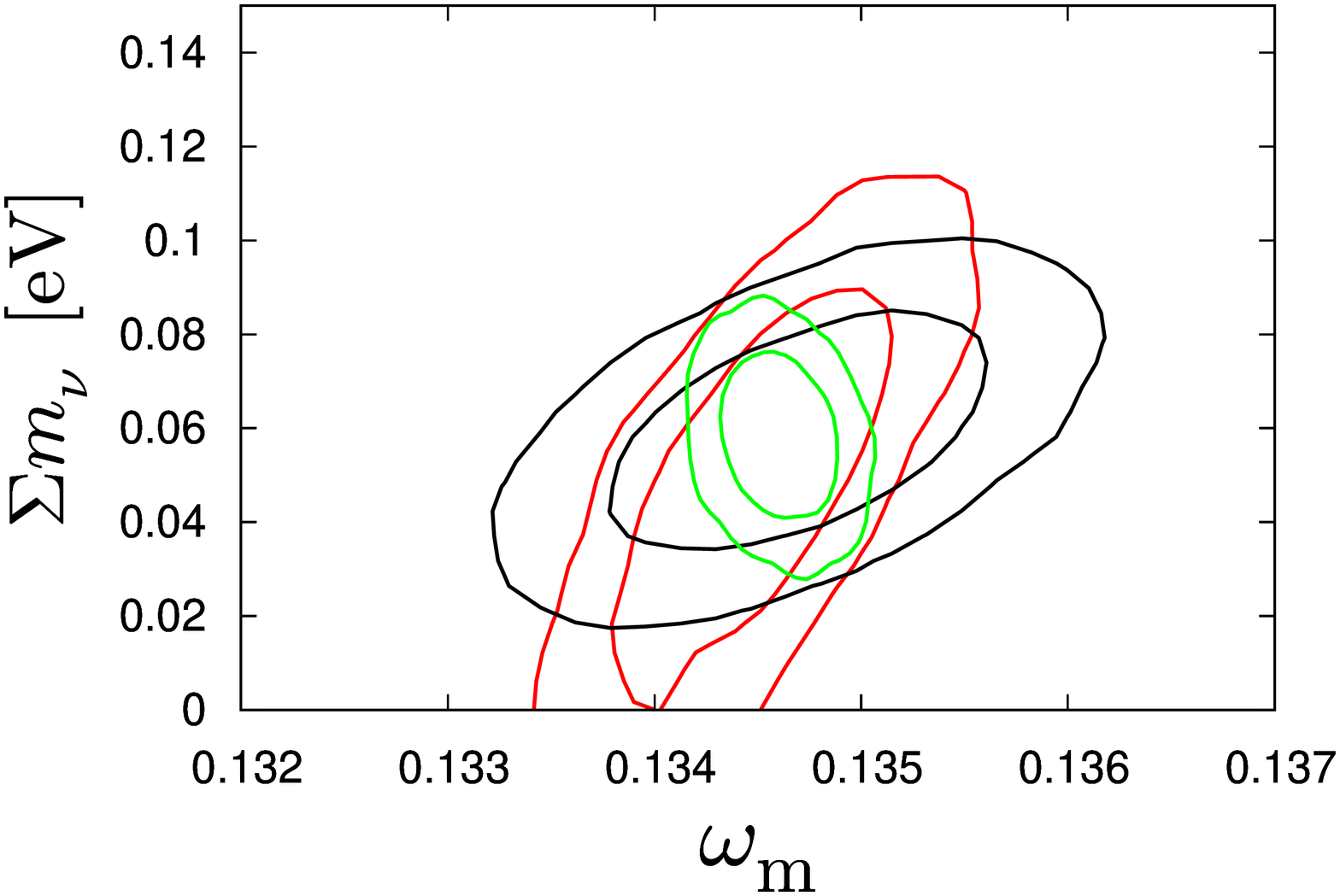}
\includegraphics[height=.48\textwidth,angle=270]{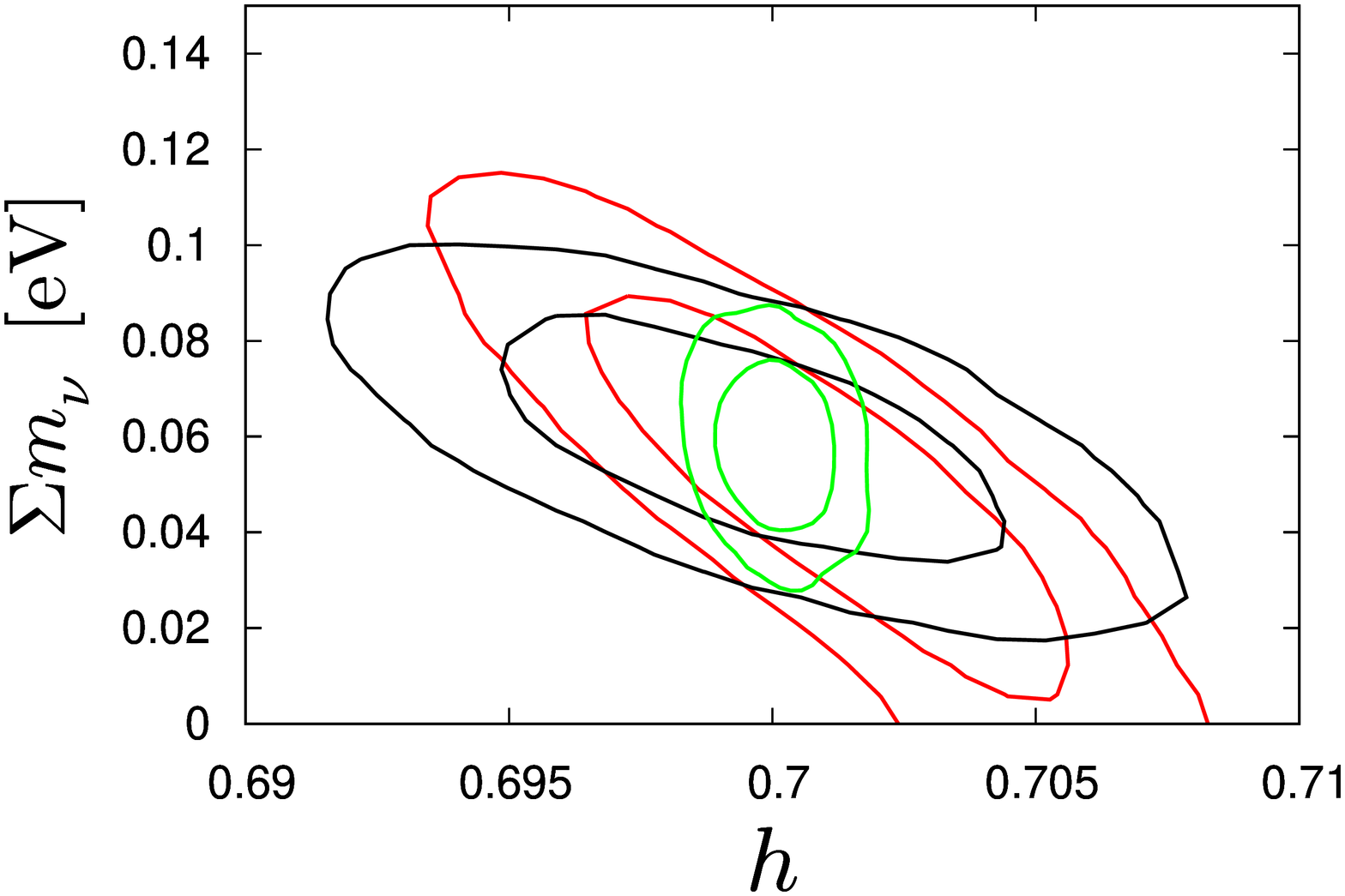}
\includegraphics[height=.48\textwidth,angle=270]{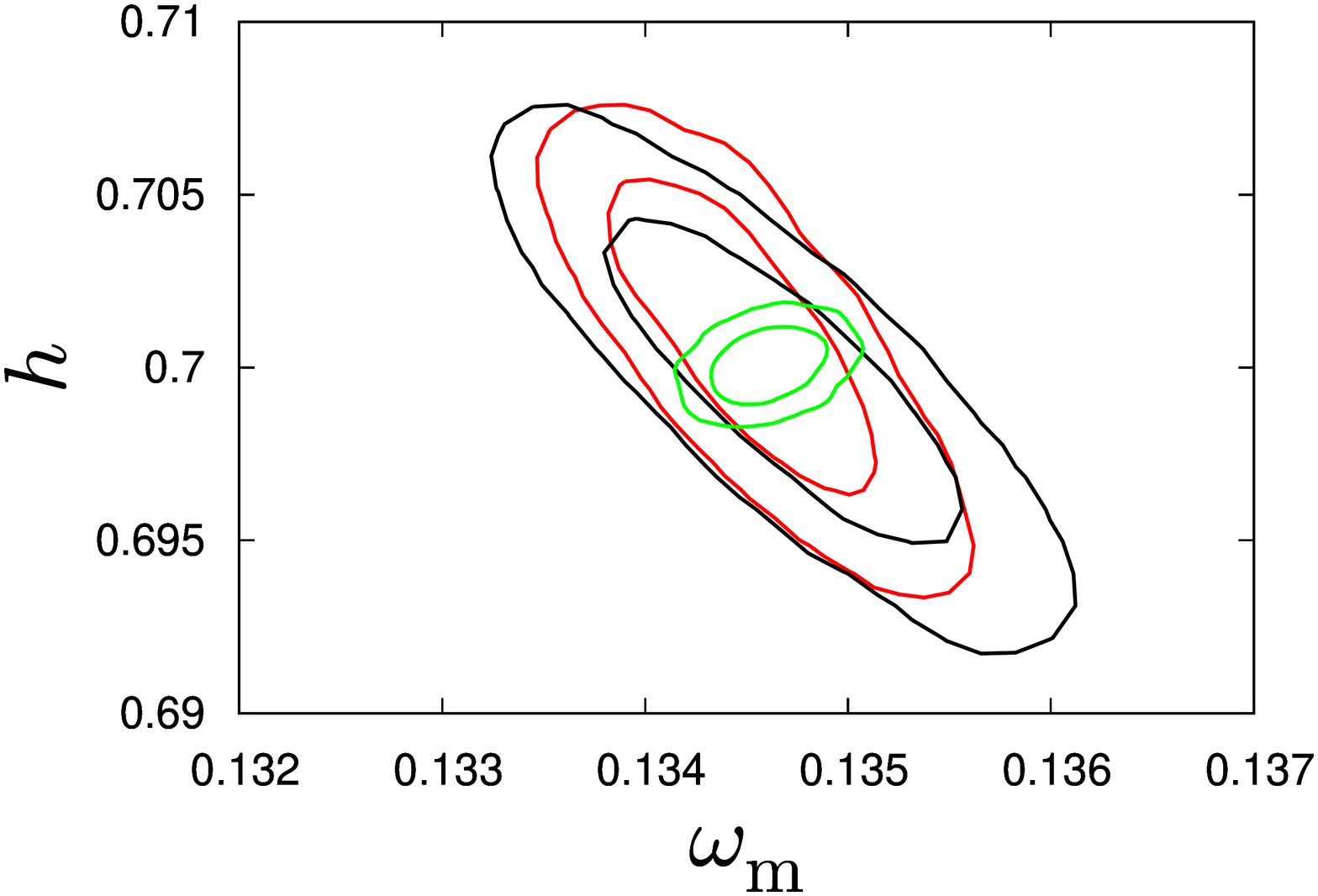}
\caption{Forecast two-dimensional joint marginalised 68\%- and 95\%-credible contours in the  $(\omega_{\rm m},h,\sum m_\nu)$-subspace.  {\it Black:}  {\sc Planck} CMB data+galaxy auto-spectrum (11~bins) without bias marginalisation. {\it Red:} {\sc Planck} CMB+shear auto-spectrum (2~bins). {\it Green:} {\sc Planck}+galaxy auto-spectrum+shear auto-spectrum+shear-galaxy cross-correlation.\label{fig:fullplot}}
\end{figure}

A first hint at an explanation for this dramatic improvement in the sensitivities
can be gleaned from examining  the joint two-dimensional 68\%- and 95\%-credible
contours in the $(\omega_{\rm m}, h, \sum m_\nu)$-subspace of our seven-dimensional model
parameter space (the other parameter directions are essentially
decoupled).  These are shown in figure~\ref{fig:fullplot}.  Already, we can
see from these two-dimensional results that the degeneracy directions in the
$(\omega_{\rm m}, h, \sum m_\nu)$-subspace are somewhat different for
{\sc Planck}+shear and for {\sc Planck}+galaxies, so that in combination they break each other's degeneracies.

\begin{figure}[t]
\center
\includegraphics[width=.48\textwidth,trim=0 0 0 0]{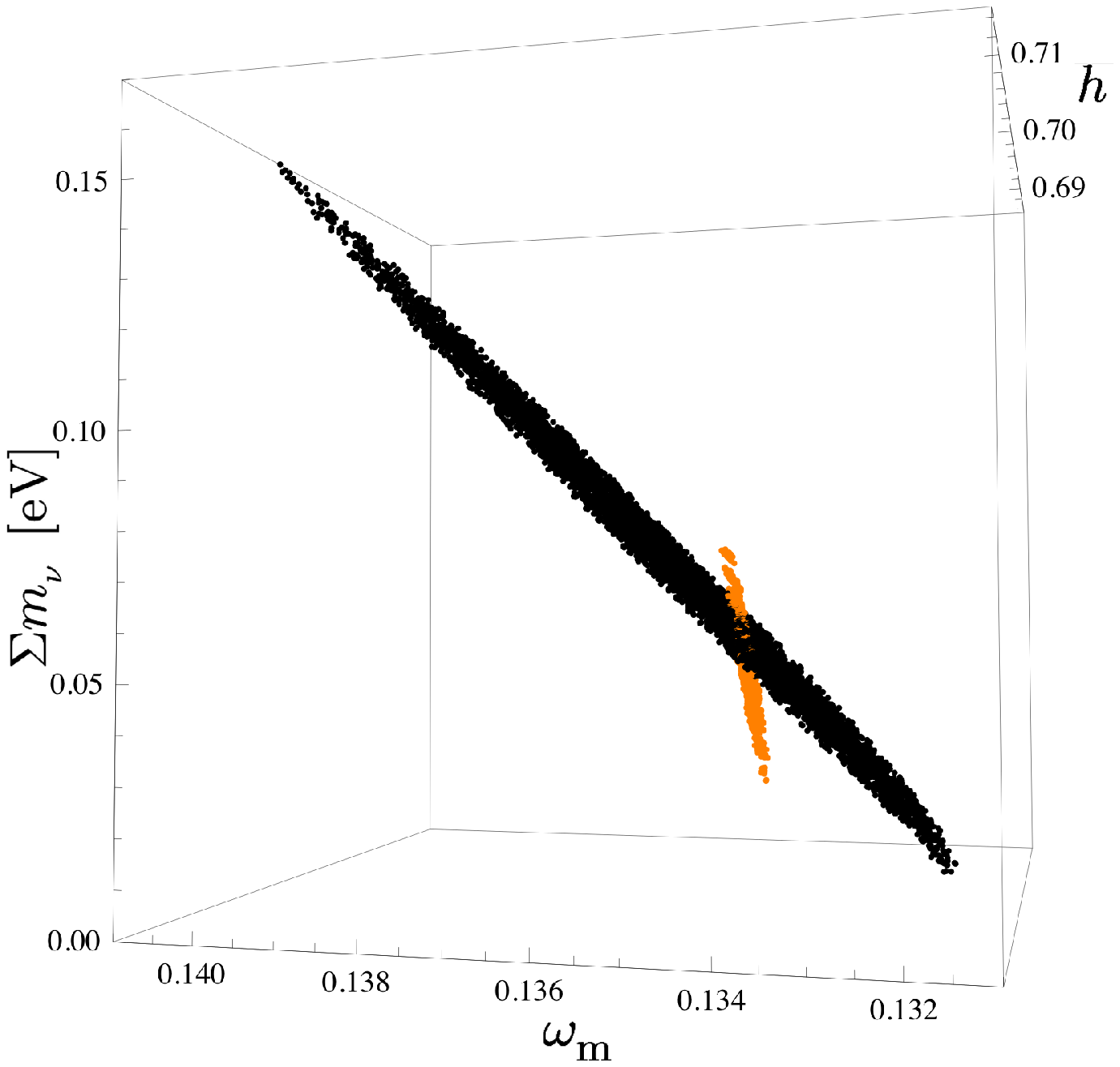}
\includegraphics[width=.48\textwidth]{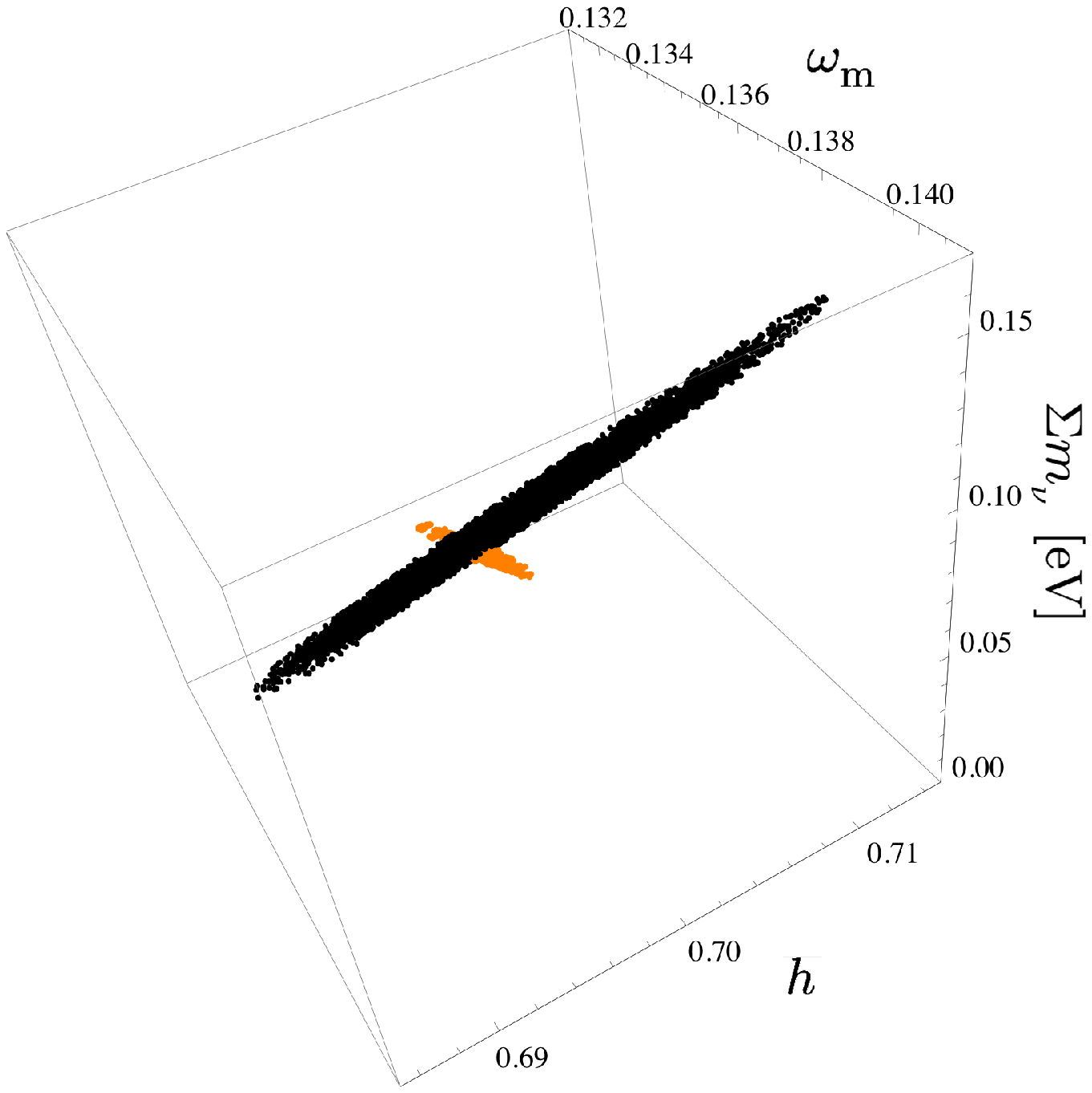}
\caption{ \label{fig:3d3pplotmain}  Three-dimensional scatter plots of samples with $\chi_{\rm eff}^2 \leq 2$ from the Markov chains of our three-parameter fits.
{\it Black}: Galaxy auto-spectrum only. {\it Orange}: {\sc Planck} CMB data+shear auto-spectrum. }
\end{figure}

A better illustration of this degeneracy-breaking can be found in
figure~\ref{fig:3d3pplotmain}.  Here, we show a three-dimensional
scatter plot of the Markov chain samples with $\chi_{\rm eff}^2 \leq
2$ from a three-parameter fit of the {\sc Planck}+shear data and the
galaxy-only data, varying in both cases only $(\omega_{\rm m}, h, \sum
m_\nu)$.  When viewed from a suitable perspective, it is clear
from the scatter plot that although the region of parameter space
preferred by the galaxy-only data is generally much larger than that
favoured by {\sc Planck}+shear, the two regions in fact have very
little overlap, and cut into each other in a way that improves
significantly the constraints on all three parameters.

For a more quantitative, semi-analytic discussion of the degeneracy
directions preferred by the different data sets in the $(\omega_{\rm
  m}, h)$- and $(\omega_{\rm m}, h, \sum m_\nu)$-subspaces, we refer
the reader to appendix~\ref{sec:paramdegeneracy}.

\subsection{The effect of shear-galaxy cross-correlation}

As we saw earlier in figure~\ref{fig:xspectra}, shear and galaxy clustering
are anti-correlated with one another.  Thus, one might reasonably expect
the addition of shear-galaxy cross-correlation to our combined analysis to yield even tighter
parameter constraints, especially because cross-correlation is not
subject to shot noise (although it is of course still subject to cosmic variance).

However, as shown in table~\ref{tab:errors}, the inclusion of shear-galaxy
cross-correlation improves the parameter uncertainties only marginally.  Comparing the ``csgx'' and the ``csg'' fits  (and also ``csg$_{\rm b}$x'' and ``csg$_{\rm b}$''),
the improvement gained from shear-galaxy cross-correlation is of order  10\%.  One reason for this modest gain is that only
for very few combinations of shear and galaxy redshift bins is the cross-correlation actually significant, e.g., low-redshift shear
will almost be uncorrelated with the high-redshift galaxy clustering.  Another reason is that over most of the range of multipoles we analyse, both the galaxy and the shear auto-spectra are already limited only by cosmic variance; there is thus little to gain from the cross-correlation being noise-free.

\subsection{The role of linear spectral shape evolution}

One distinctive signature of massive neutrinos is the scale-dependence of the linear growth function $G^{\rm lin}(k,z)$ induced by neutrino free-streaming.  This scale-dependence means that $k$-modes generally grow at the different rates, so that the shape of the linear 3D matter power spectrum, especially the degree of power suppression at large $k$  due to
non-clustering neutrinos, effectively changes with time.  However, this spectral shape evolution is typically a much smaller effect than the actual free-streaming suppression itself~\cite{Hannestad:2007cp}, and until we can measure the suppression with high significance, it will most likely remain unobservable.  To this end, in view of the extraordinary sensitivity of a {\sc Euclid}-like survey to the neutrino mass, we would like to reexamine  the detectability of linear spectral shape evolution in this section.

We consider three fiducial neutrino mass sums $\sum m_\nu$: (i) the default value of 0.058~eV used so far in our analysis, (ii) 0.11~eV, corresponding to the minimum mass of the inverted hierarchy (see also section~\ref{sec:hierarchy}), and (iii) 0.465~eV, which is roughly the 95\% upper limit on $\sum m_\nu$ from
 the current generation of cosmological probes.    For each of these fiducial models we generate mock data sets as per usual, but fit them using
 a modified linear transfer function that has the  {\it shape} of the $z=0$ linear transfer function and an {\it amplitude} scaled by the $k \to 0$ linear growth functions, i.e.,
 \begin{equation}
 \label{eq:modT}
 T_{\Psi,\delta}^{\rm lin,mod}(k,z) =  \frac{G^{\rm lin}(k\to 0,z)}{G^{\rm lin}(k \to 0,0)} T^{\rm lin}_{\Psi,\delta}(k,0) = \frac{T^{\rm lin}_{\Psi,\delta}(k\to 0,z)}{T^{\rm lin}_{\Psi,\delta}(k \to 0,0)} T^{\rm lin}_{\Psi,\delta}(k,0) .
  \end{equation}
 Nonlinear corrections (which also induce a scale-dependence in the effective growth functions) are computed with {\sc Halofit}  using this modified linear transfer function as an input.  If linear spectral shape evolution does indeed constitute an observable effect, it will manifest itself in the fit as a bias in the parameter estimates and/or a degradation of the best-fit $\chi_{\rm eff}^2$.

\begin{figure}[t]
\center
\includegraphics[width=.7\textwidth,angle=270]{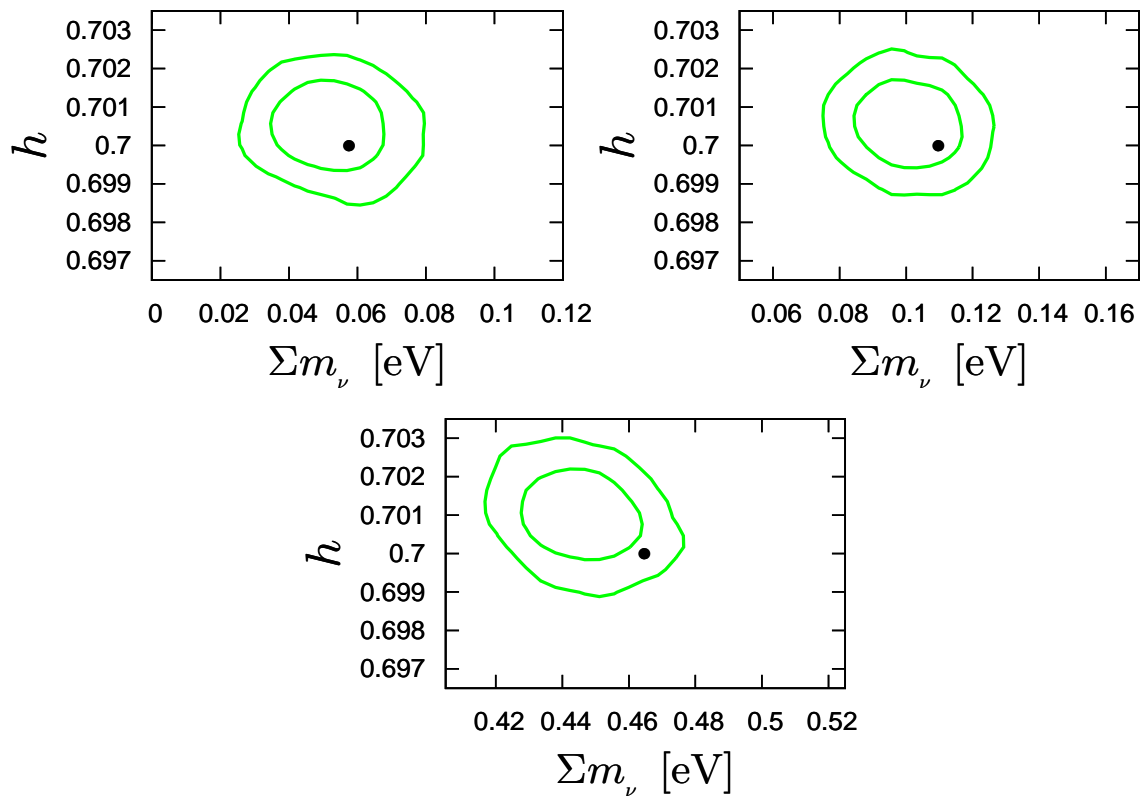}
\caption{\label{fig:shape} Two-dimensional joint marginalised 68\%- and 95\%-credible contours in the  $(h,\sum m_\nu)$-subspace from a ``csgx'' fit, where the analysis has been performed using the modified linear transfer function~(\ref{eq:modT}) which has no shape evolution.  Black dots mark the fiducial values of $\sum m_\nu$ and $h$ in each model.}
\end{figure}

Figure~\ref{fig:shape} shows the two-dimensional joint marginalised 68\%- and 95\%-credible contours in the $(h,\sum m_\nu)$-subspace from this exercise, computed using the most stringent  ``csgx'' data set.
These two parameters suffer the strongest bias when linear spectral shape evolution has been deliberately ignored.  Nonetheless, unless $\sum m_\nu$ is close to or beyond its present-day upper limit,
the deviation induced is considerably below $1 \sigma$; even in the case of \mbox{$\sum m_\nu^{\rm fid}=0.465$}, we do not expect the parameter bias to exceed $\sim 1.5 \sigma$.

A somewhat stronger effect can been seen from an examination of each model's  goodness-of-fit relative to that of the true underlying model.  For the three scenarios considered here, we find, in ascending order of fiducial neutrino masses,   $\Delta \chi_{\rm eff}^2=1.92, 2.89, 7.68$.  Thus, in the very optimistic case of a true $\sum m_\nu$ that is as large as is presently allowed by observations, a model comparison analysis may yield some mild evidence for the time evolution of the linear spectral shape. These numbers also indicate that spectral evolution due to neutrino free-streaming plays but a negligible role in the  constraint of the neutrino mass by a {\sc Euclid}-like survey.

\subsection{Modelling of the neutrino mass spectrum\label{sec:hierarchy}}

In our analysis so far we have always assumed one massive and 2.046 massless neutrinos.  This mass spectrum is clearly inconsistent with both the normal (NH) and the inverted (IH) hierarchies stipulated by neutrino oscillation experimental data:
\begin{align}
m_1 &\quad <  \quad m_2 = m_1 + \sqrt{\Delta m_{12}^2}   \quad <  \quad m_3 = m_1+  \sqrt{\Delta m_{12}^2} + \sqrt{\Delta m_{23}^2}, \qquad {\rm NH}, \\
m_3 & \quad  < \quad m_1 = m_3 + \sqrt{\Delta m_{23}^2}-\sqrt{\Delta m_{12}^2}  \quad < \quad  m_2 = m_3 + \sqrt{\Delta m_{23}^2}, \qquad {\rm IH},
\label{eq:inverted}
\end{align}
where $\Delta m_{12}^2 = 7.5 \times 10^{-5} \, {\rm eV}^2$ and $|\Delta m_{23}^2| = 2.3 \times 10^{-3} \, {\rm eV}^2$~\cite{Fogli:2012ua,Tortola:2012te}.  In view of the remarkable forecast sensitivities to $\sum m_\nu$ presented in table~\ref{tab:errors}, one might wonder if our simple ``1~massive+2~massless'' scheme will remain an adequate approximation in the face of real {\sc Euclid} data.

The question of whether the neutrino mass spectrum could ever be resolved by precision cosmology has been discussed previously in many different contexts (see, e.g., \cite{Lesgourgues:2004ps,Jimenez:2010ev}).   Here, we wish to establish (i) if using an approximate model would bias the parameter inference (in $\sum m_\nu$ or other parameters), and (ii) if an approximate model would provide a much worse fit to the data than the true model, so that the former could be excluded by way of model comparison.

Our set-up is simple.  Assuming a fiducial sum of $\sum m_\nu = 0.11$~eV, we generate two mock data sets,  one with three non-degenerate neutrino masses distributed according to the normal hierarchy, and the second distributed according to the inverted hierarchy.  Each of these two sets is analysed using three different approximate models for the neutrino mass spectrum:
\begin{itemize}
\item ``1+2'': 1 massive state, 2.046 massless states
\item ``2+1'': 2 degenerate massive states, 1.046 massless state
\item ``3+0'': 3 degenerate massive states, 0.046 massless state
\end{itemize}
The choice of  $\sum m_\nu =0.11$~eV as the fiducial sum is motivated by the fact that it is roughly the minimum value allowed by the inverted hierarchy according to equation~(\ref{eq:inverted}).

\begin{figure}[t]
\center
\includegraphics[height=.48\textwidth,angle=270]{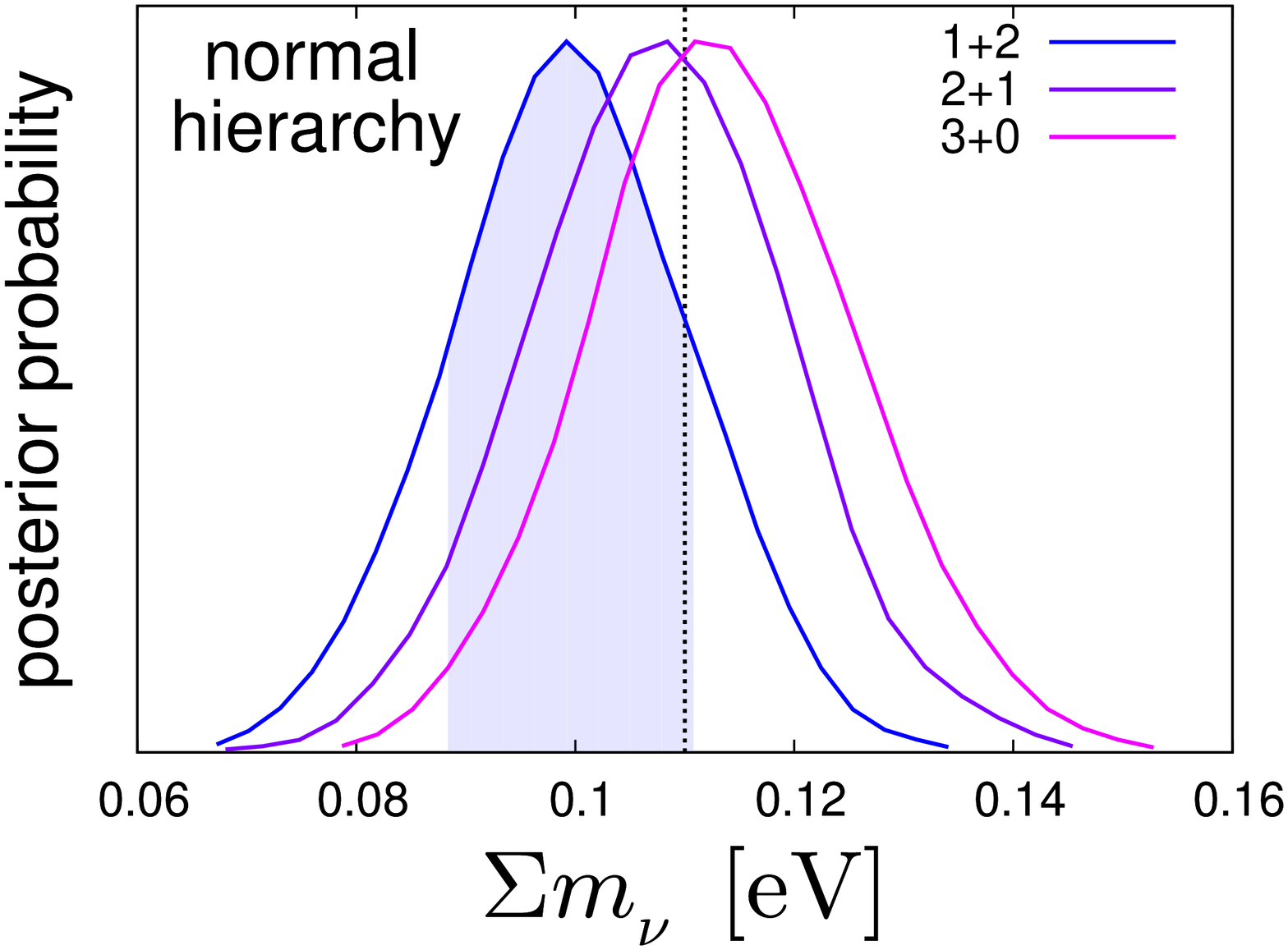}
\includegraphics[height=.48\textwidth,angle=270]{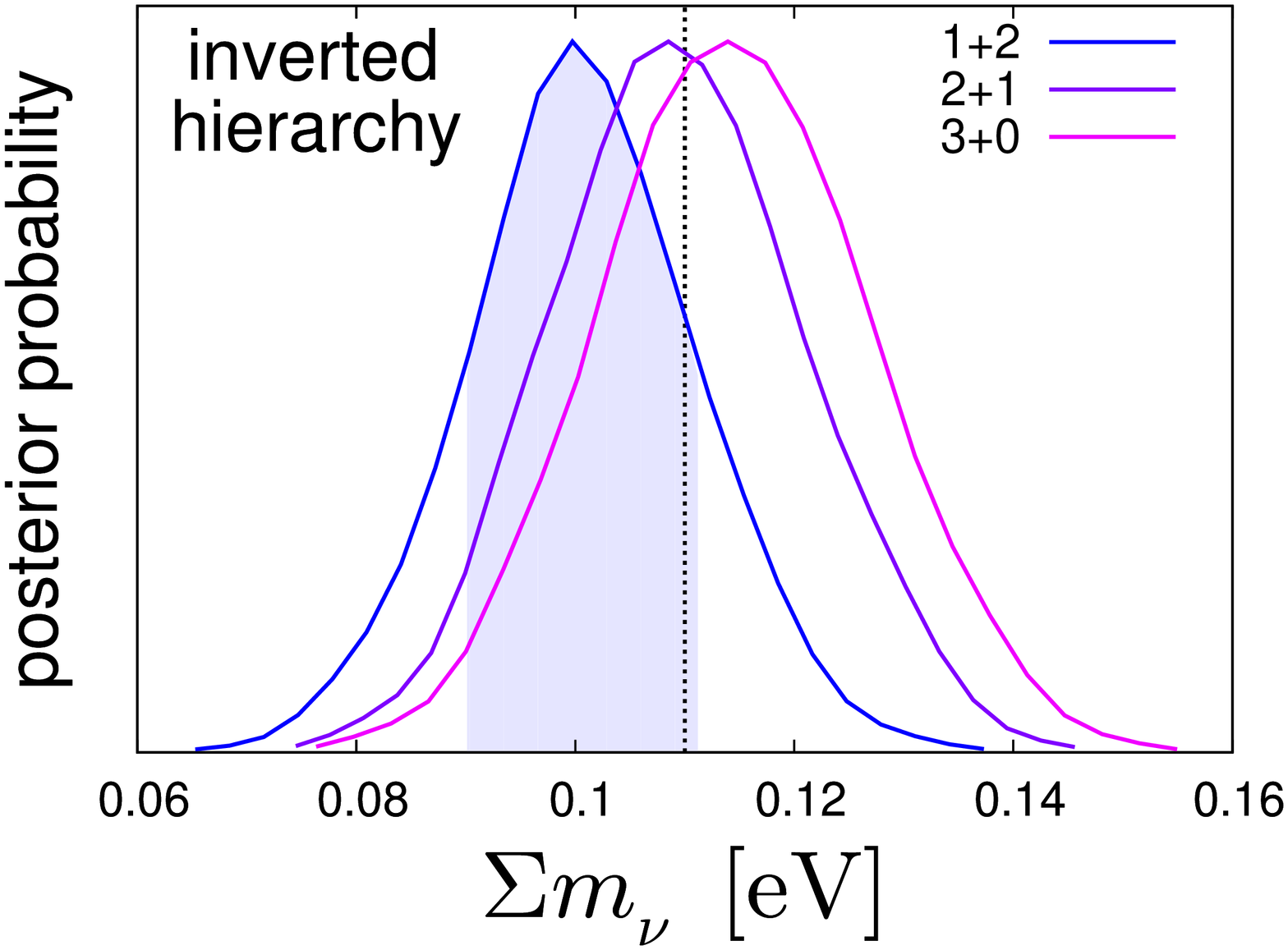}
\caption{Marginalised posterior probability density for the sum of neutrino masses from a ``csgx'' fit, where the analysis has been performed in the 1+2, 2+1, and 3+0 schemes.  The fiducial model has $\sum m_\nu = 0.11$~eV (marked by the dotted vertical line) and assumes the normal ({\it left}) and the inverted ({\it right}) hierarchy, respectively.  The shaded area indicates the minimal 68\%-credible interval for the worst-fitting scenario.\label{fig:hierarchyplots}}
\end{figure}

Figure~\ref{fig:hierarchyplots} shows the one-dimensional marginalised posteriors for $\sum m_\nu$ from our six fits, using a combination of {\sc Planck} CMB, galaxy (no bias marginalisation), shear, and shear-galaxy cross-correlation data (``csgx'', using the notation of table~\ref{tab:errors}).  Expectably, of the seven free parameters in our model,  the estimation of $\sum m_\nu$ is  most strongly affected by how
we choose to model the neutrino mass spectrum.  Even so, the bias in the inference of $\sum m_\nu$ caused by a bad model choice is at worst a $1\sigma$ effect, one that we can easily live with.  Likewise, in all cases, the goodness-of-fit as quantified by the model's best-fit $\chi_{\rm eff}^2$ suffers only a  degradation of at most
$\Delta \chi_{\rm eff}^2 = 0.9$ relative to the true model, indicating that the data show no strong preference for any particular neutrino mass spectrum.
We therefore conclude that even under the best circumstances, the combination of {\sc Planck} and {\sc Euclid} data will not be able to probe directly the neutrino mass hierarchy, and hence, an exact modelling of the neutrino mass spectrum for the purpose of cosmological parameter inference is  unnecessary.

Finally, we observe in figure~\ref{fig:hierarchyplots} that for both the normal and the inverted hierarchy, the 2+1 scheme appears to be the most unbiased approximation.  While this makes perfect sense for the inverted hierarchy given the fiducial sum of $\sum m_\nu=0.11$~eV, we note that in the normal hierarchy, this fiducial model in fact
 corresponds to one state with a 0.05~eV mass, and two almost degenerate, slightly lighter states at 0.03~eV.
The best approximate mass spectrum should therefore lie somewhere between the 1+2 and the 3+0 models.  The 2+1 scheme happens to satisfy this ``in-between'' requirement and hence provide an almost unbiased fit to the mock data, although strictly speaking it is a very poor approximation to the underlying physics.

\section{More complex cosmological models\label{sec:complex}}

So far we have derived our constraints on neutrino masses and other cosmological parameters in the context of a simple 7-parameter model, namely, a $\Lambda$CDM cosmology extended with one additional parameter  $\sum m_\nu$.  However, previous studies have shown that $\sum m_\nu$, as well as the physical matter density $\omega_{\rm}$ and Hubble parameter $h$, can be highly  degenerate with physics not encompassed by this simple 7-parameter model, for instance, a nonstandard value for the effective number of massless neutrino species $N_{\rm eff}^{\rm ml}$, or a dark energy equation of state parameter $w$  that deviates from $-1$ (see, e.g,~\cite{Hannestad:2005gj,Hannestad:2003ye,Hamann:2010bk,Wang:2012vh,RiemerSorensen:2012ve}).   It is therefore interesting to ask if the extremely tight constraints on  $\omega_{\rm m}$, $h$ and $\sum m_\nu$
reported in section~\ref{sec:results} would survive in a more complex model parameter space.

\begin{figure}[t]
\center
\includegraphics[height=.88\textwidth, angle=270]{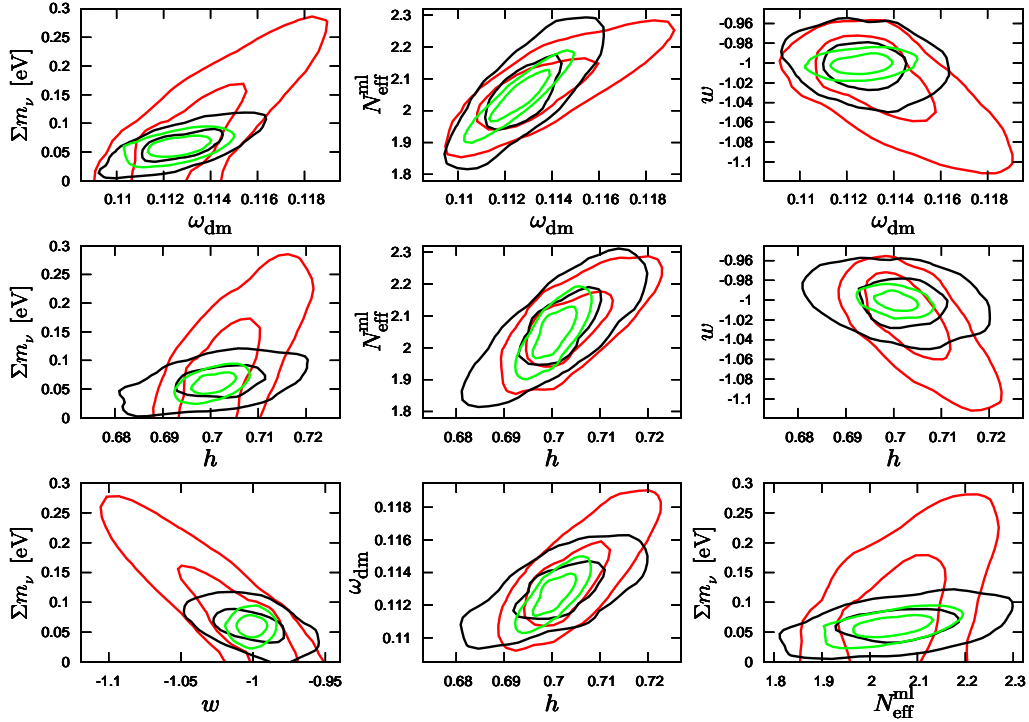}
\caption{Forecast two-dimensional joint marginalised 68\%- and 95\%-credible contours in the $(\omega_{\rm m}, h, \sum m_\nu, N_{\rm eff}^{\rm ml},w)$-subspace of the 9-parameter model (equation~(\ref{eq:9param})).
{\it Black:}  {\sc Planck} CMB data+galaxy auto-spectrum (11~bins) without bias marginalisation. {\it Red:} {\sc Planck} CMB+shear auto-spectrum (2~bins). {\it Green:} {\sc Planck}+galaxy auto-spectrum+shear auto-spectrum+shear-galaxy cross-correlation.
\label{fig:complex}}
\end{figure}

To test this, we perform a forecast  for a 9-parameter model described by
\begin{equation}
\Theta_9 = (\omega_{\rm b},\omega_{\rm dm},h,A_{\rm s},n_{\rm s},z_{\rm re},f_\nu,N_{\rm eff}^{\rm ml},w),
\label{eq:9param}
\end{equation}
using the ``csgx'' data set.  We choose the fiducial values
\begin{equation}
\Theta_{9, {\rm fid}} = (0.022, 0.1126228, 0.7, 2.1 \times 10^{-9}, 0.96, 11, 0.00553, 2.046, -1),
\end{equation}
i.e., the same fiducial values as  those used earlier in our 7-parameter fit (see equation~(\ref{eq:fidvalues})),
with the standard model prediction for $N_{\rm eff}^{\rm ml}$ and dark energy in the form of a cosmological constant. Note that we assume, as we did before, one massive neutrino species;  only the number of massless species is now free to vary.

Figure~\ref{fig:complex} shows the correlations between $(\sum m_\nu, \omega_{\rm m}, h)$ and the two new parameters.
A strong three-way degeneracy persists between $(\omega_{\rm m}, h, N_{\rm eff}^{\rm ml})$, since these are the main parameters that control the epoch of matter-radiation equality and the CMB angular sound horizon%
\footnote{The sound horizon depends also on the baryon density $\omega_{\rm b}$.  However, $\omega_{\rm b}$ can be very well constrained by the CMB even-to-odd peak ratios, and is therefore effectively decoupled from the degeneracy problem.}
 (see section~II of~\cite{Abazajian:2012ys} for an explanation).  This three-way degeneracy can lead to an even stronger apparent degeneracy between $(\omega_{\rm m},h,\sum m_\nu)$ (and, by association, between $m_\nu$ and $N_{\rm eff}^{\rm ml}$) in the matter power spectrum,
 although the degeneracy is not an exact one and  once shear and galaxy data are used in combination,
only a small correlation remains between these parameters.
  However, while the $(\omega_m,h,N_{\rm eff}^{\rm ml})$-degeneracy does not translate into a significant degradation in the sensitivity to the neutrino mass,  it does nonetheless turn the mildly negative correlation between $\omega_m$ and $\sum m_\nu$ previously seen in the 7-parameter fit into a positive one.  The $(\sum m_\nu,w)$-degeneracy, on the other hand, is completely broken because of the combination of two independent determinations of the growth function $G(k,z)$, which tightly limits the $(\omega_{\rm dm},h,w)$-degeneracy inherent in $G(k,z)$.

 \begin{table*}[t]
  \caption{Posterior standard deviations for the parameters $\omega_{\rm dm}$, $h$, $\sum m_\nu$, $N_{\rm eff}^{\rm ml}$, and $w$ in the 9-parameter model, derived from the ``cs'', ``cg'', ``csgx'' and ``csg$_{\rm b}$x'' data sets.  For easier comparison we tabulate also the corresponding values from the ``csgx'' and ``csg$_{\rm b}$x'' sets in the 7-parameter model.\label{tab:7vs9}}
\begin{center}
{\footnotesize
  \hspace*{0.0cm}\begin{tabular}
  {lccccc} \hline \hline
  Data & $10^3 \times \sigma(\omega_{\rm dm})$ & $100 \times \sigma(h)$ & $\sigma(\sum m_\nu)/$eV & $\sigma(N_{\rm eff}^{\rm ml})$ & $\sigma(w)$ \\       \hline
cs &   1.864 & 0.638 & 0.064 & 0.081 & 0.0310 \\
cg &   1.121 & 0.655 & 0.020 & 0.086 & 0.0163 \\
csg & 0.885 & 0.324 & 0.012 & 0.056 & 0.0083 \\
csgx & 0.874 & 0.292 & 0.012 & 0.055 & 0.0065 \\
csg$_{\rm b}$ & 1.400 & 0.529 & 0.042 & 0.068 & 0.0207 \\
csg$_{\rm b}$x & 1.390 & 0.482 & 0.042 & 0.068 & 0.0186 \\
\hline
csgx (7-parameter)& 0.181 & 0.071 & 0.011 & - & - \\
csg$_{\rm b}$x (7-parameter) & 0.354 & 0.244 & 0.022&- & - \\
\hline \hline
  \end{tabular}
  }
  \end{center}
\end{table*}

To check that it is indeed the combination of galaxy and shear data that saves the bound on $\sum m_\nu$ in the extended model, we perform two separate forecasts using
{\sc Planck}+galaxy (``cg'') and  {\sc Planck}+shear (``cs'') respectively, the results of which are shown also in figure~\ref{fig:complex}.   Clearly,
removing either the galaxy or the shear data from the combined fit restores the strong degeneracy between $\omega_m$ and $\sum m_\nu$, and hence the
degeneracies of $\sum m_\nu$ with $w$ and~$N_{\rm eff}^{\rm ml}$.  The same trend can also been seen in table~\ref{tab:7vs9}, where we list the posterior standard deviations from various cosmological parameters.
In the 9-parameter model, the sensitivities to $\sum m_\nu$ and $w$ degrade both by a factor of five when galaxy data are removed, and by a factor of two and three, respectively,
when shear data are discarded.  This is in contrast with the 7-parameter model, where the corresponding degradations in the $\sum m$ sensitivity are 200\% and 50\% respectively according to table~{\ref{tab:errors}.
Thus, we conclude that  the simple combination of galaxy and shear data is even more powerful for constraining the extended model parameter space than for
 the minimal 7-parameter model.  Shear-galaxy cross-correlation, on the other hand, continues to play a marginal role in the parameter constraints; excluding ``x'' from the analysis deteriorates the sensitivities by no more than 10\%.

Lastly, we examine the impact of galaxy bias marginalisation on the parameter sensitivities in the 9-parameter model.  Full marginalisation over an unknown linear galaxy bias  means a complete loss of information on the normalisation and hence the growth function $G(k,z)$ from the galaxy data, the result of which is a severe degradation in parameter sensitivities to $(\omega_{\rm dm},h,w)$ because of their mutual degeneracy in $G(k,z)$.  As shown in table~\ref{tab:7vs9}, the sensitivities of the ``csg$_{\rm b}$x'' data set to
$(\omega_{\rm dm},h,w)$ degrade by a factor 1.5 to 3 relative to the numbers from ``csgx''.  In contrast, the sensitivity to $N_{\rm eff}^{\rm ml}$ suffers only a 20\% setback,
understandably because $N_{\rm eff}^{\rm ml}$ does not play a role in the late-time normalisation of matter power spectrum.
For $\sum m_\nu$, the unbroken $(\omega_{\rm dm},h,w)$-degeneracy translates into a degradation of the $1\sigma$ sensitivity to $0.042$~eV, and reduces the detection significance to only $1.5\sigma$ should the true $\sum m_\nu$ be at its most pessimistic value of 0.06~eV.   Shear-galaxy cross-correlation plays essentially no role in any of these constraints.
Nonetheless, we note that compared with using shear data alone (``cs''), the addition of unnormalised galaxy data to the analysis still offers a 20\% to 30\% improvement in most parameter sensitivities; shape information still contributes some constraining power.

\section{Conclusions\label{sec:conclusions}}

In this work, we have carefully studied the capacity of a future photometric galaxy survey such as {\sc Euclid} in terms of its constraining power on the neutrino mass.
Several previous studies have also considered the idea of using either the cosmic shear or the galaxy clustering power spectrum from a {\sc Euclid}-like survey in conjunction with CMB measurements from the {\sc Planck} mission to probe the neutrino mass scale
(e.g., \cite{Hannestad:2006as,Jimenez:2010ev,Kitching:2008dp,Namikawa:2010re,Joudaki:2011nw,2007MNRAS.381.1313A}).
Indeed, as a first test, we have confirmed that {\sc Planck}+shear and {\sc Planck}+galaxy power spectrum are {\it individually} sensitive to the neutrino mass sum $\sum m_\nu$ at the $\sigma(\sum m_\nu) = 0.02 \to 0.04$~eV level depending on our knowledge of the linear galaxy bias, compatible with earlier findings.

An interesting interplay arises, however, when shear and galaxy clustering data are used in {\it combination}.  Here, as we demonstrate in a new study, the parameter degeneracy directions in the shear and the galaxy data in the three-dimensional $(\omega_{\rm m},h,\sum m_\nu)$-subspace are fundamentally very different.  This means that the two data sets, although seemingly measuring the same thing, are in fact complementary to one another, and when used in combination, conspire to lift each other's parameter degeneracies.  For the neutrino mass measurement, this complementarity between the shear and the galaxy power spectra can lead to a significant improvement in the survey's sensitivity to $\sum m_\nu$: in the case of complete knowledge of the linear galaxy bias, the sensitivity can potentially reach $\sigma (\sum m_\nu) =0.011$~eV, which will constitute a $5.4\sigma$ detection of $\sum m_\nu$ even if the true $\sum m_\nu$ has the lowest value of $\sim 0.06$~eV allowed by neutrino oscillation experiments in the normal hierarchy.  In the opposite extreme, however, where no knowledge of the bias is available,
 the improvement anticipated in $\sigma (\sum m_\nu)$ from combining shear and galaxy measurements is greatly diminished, mainly because the galaxy data no longer contain any useful information on  the amplitude of the matter power spectrum.  Nonetheless, a sensitivity of $\sigma (\sum m_\nu) = 0.022$~eV, or a detection significance of  $2.5\sigma$, can still be expected.

We have also verified explicitly that a similar sensitivity of $\sigma(\sum m_\nu) = 0.012$~eV can be attained in a more complex cosmological model extended with two additional free parameters, the effective number of relativistic degrees of freedom $N_{\rm eff}^{\rm ml}$ and the dark energy equation of state parameter $w$, if the linear galaxy bias is exactly known.  In the complete absence of this knowledge, however, the sensitivity to $\sum m_\nu$ deteriorates to $0.042$~eV, thereby reducing the detection significance to $1.5\sigma$ should the true $\sum m_\nu$ be a pessimistic 0.06~eV.

Given these spectacular sensitivities, we have also examined if a {\sc Euclid}-like survey might be able to resolve the neutrino mass spectrum, and hence distinguish between the normal and the inverted hierarchy in a direct way. Taking a fiducial cosmological model with $\sum m_\nu = 0.11$ eV and three non-degenerate neutrino masses distributed according to the normal or the inverted hierarchy,  we find that parameter estimation, even under the most aggressive model and data assumptions,
suffers no significant bias (less than $1\sigma$) when a grossly incorrect neutrino mass spectrum has been assumed in the statistical inference.  The same is true also when we consider the goodness-of-fit of the wrong models; the degradation in $\chi_{\rm eff}^2$ is always less than one unit relative to the true model.

Thus, we conclude that  although a survey like  {\sc Euclid} will very likely provide a positive detection of neutrino mass at high significance, the exact nature of the neutrino mass spectrum remains out of its reach, unless the total neutrino mass is sufficiently low to rule out the inverted hierarchy.

\section*{Acknowledgements}
We acknowledge computing resources from the Danish Center for
Scientific Computing (DCSC).


\appendix

\section{Parameter degeneracies\label{sec:paramdegeneracy}}

In this appendix we examine more closely the degeneracies amongst the parameters $\omega_{\rm m}$, $h$, and $\sum m_\nu$ in the context of restricted two-parameter and three-parameter fits wherein only these parameters or a subset of them are let to vary.  As already discussed in section~\ref{sec:binning}, these three are the only cosmological parameters
in our seven-parameter model~(\ref{eq:model}) for which the addition of galaxy and/or shear auto-spectrum measurements to {\sc Planck} CMB data is expected to improve significantly their uncertainties.

We begin with a two-parameter analysis varying only $\omega_{\rm m}$ and $h$, after which we extend to model space to include also a free $\sum m_\nu$.

\subsection{Two-parameter fits\label{sec:twoparameters}}

\subsubsection{CMB only}

The most well-measured feature in the CMB temperature anisotropies is the angular sound horizon $\theta_{\rm s} \equiv r_{\rm s}(z_\star)/D_A(0,z_\star)$.
In the numerator, $r_{\rm s}(z_\star)$ is the physical sound horizon at decoupling $z_\star \simeq 1100$,
\begin{eqnarray}
r_{\rm s}(z_\star) & \equiv & \int _0^{\tau(z_\star)} \di \tau \ c_{\rm s}(\tau) \nonumber \\
& \simeq & 3000 \sqrt{\frac{4}{3} \frac{1}{\omega_{\rm m}} \frac{a(z_{\rm eq})}{R(z_{\rm eq})} }
\ln \left[\frac{ \sqrt{1+R(z_\star)}+\sqrt{R(z_\star) +R(z_{\rm eq})}}{1+ \sqrt{R(z_{\rm eq})}}\right]  {\rm Mpc},
\end{eqnarray}
where $R(z) = (3/4) (\omega_{\rm b}/\omega_\gamma) a(z)$ is the baryon-to-photon ratio, and
$1+ z_{\rm eq} =  (\omega_{\rm m}/\omega_\gamma) (1 + 0.227 N_\nu)^{-1}$,
with $\omega_\gamma = 2.4692 \times 10^{-5}$, gives the redshift of matter-radiation equality. In the denominator,
\begin{equation}
D_A(0,z_\star) \equiv \int_{0}^{z_\star} \frac{\di z}{H(z)} \simeq 3000  \int_{0}^{z_\star} \frac{\di z}{ \sqrt{\omega_{\rm m} (1+z)^3 + (h^2-\omega_{\rm m})}} \ {\rm Mpc}
\end{equation}
denotes the comoving angular diameter distance to the last scattering surface under the assumption of a flat spatial geometry.

We adopt fixed values for $\omega_{\rm b} = 0.022$ and  $N_\nu =3.046$, and vary $\omega_{\rm m}$ and $h$ around their fiducial values of $\omega_{\rm m}^{\rm fid} = 0.1346228$ and $h^{\rm fid}=0.7$.
Assuming  $\theta_{\rm s}$ can be parameterised as $\theta_{\rm s} = \omega_{\rm m}^\mu h^\nu$
in the vicinity of $(\omega_{\rm m}^{\rm fid},h^{\rm fid})$, perturbing around this point gives
\begin{equation}
\label{eq:dthetas}
\frac{\di \theta_{\rm s}}{\theta_{\rm s}} = \mu \frac{\di \omega_{\rm m}}{\omega_{\rm m}} + \nu \frac{\di h}{h},
\end{equation}
with coefficients
\begin{align}
\mu &= \left [ \frac{\partial \log r_{\rm s}}{\partial \log \omega_{\rm m}} - \frac{\partial \log D_A}{\partial \log \omega_{\rm m}} \right]_{h=h^{\rm fid},\omega_{\rm m} = \omega_{\rm m}^{\rm fid}} \simeq 0.16,  \\
\nu &= \left. - \frac{\partial \log D_A}{\partial \log h}\right|_{h=h^{\rm fid},\omega_m = \omega_{\rm m}^{\rm fid}} \simeq 0.20.
\end{align}
By demanding $\di \theta_{\rm s} =0$, equation~(\ref{eq:dthetas}) can be rearranged to give
$h - h^{\rm fid}=M (\omega_{\rm m}-\omega_{\rm m}^{\rm fid})$,
where
\begin{equation}
\label{eq:mslope}
M = - \frac{\mu}{\nu} \frac{h^{\rm fid}}{\omega_{\rm m}^{\rm fid}}\simeq -4.2
\end{equation}
gives the slope of the degeneracy ellipse in the $(x=\omega_{\rm m},y=h)$-plane.

Figure~\ref{fig:2pplot} shows the posterior in the $(\omega_{\rm m},h)$-plane for various mock data sets.   The degeneracy slope predicted in equation~(\ref{eq:mslope}) matches exactly the degeneracy direction preferred by the mock  {\sc Planck} CMB data.

\begin{figure}[t]
\center
\includegraphics[height=.68\textwidth,angle=270]{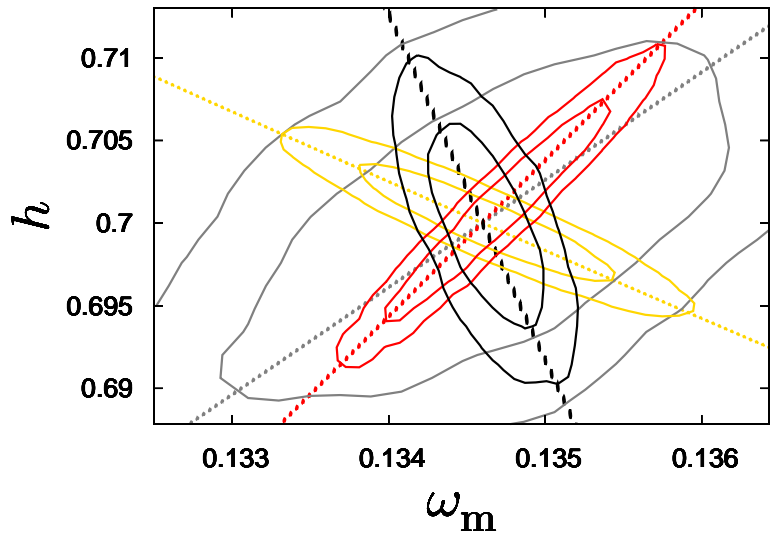}
\caption{Approximate principal degeneracy directions and forecast 68\%- and 95\%-credible contours for $h$ and $\omega_{\rm m}$ for the two-parameter model.  All other cosmological parameters are kept fixed at their fiducial values.  {\it Black:} Galaxy auto-spectrum (8~bins) without bias marginalisation.  The straight line indicating the degeneracy direction has slope $M=-21$.
{\it Grey:} Galaxy auto-spectrum data (8~bins) with bias marginalisation.  The corresponding slope is  $M=6.5$.  {\it Red:} Shear auto-spectrum data (8~bins), with $M=9.5$.   {\it Gold:} {\sc Planck} CMB data,  with $M = -4.2$. \label{fig:2pplot}}
\end{figure}

\subsubsection{Galaxy angular power spectrum}

\paragraph{Shape information only}  Full  marginalisation  over the linear bias
as discussed in section~\ref{sec:bias} forces us to discard all normalisation information, including information on the linear growth of structures as a function of redshift.
In this case, the only handle by which cosmological parameter estimation is possible is through the shape of the angular power spectrum, notably through the ``shape parameter''~$k_{\rm eq} \equiv a(z_{\rm eq}) H(z_{\rm eq})$, which determines primarily the location of the turning point of the 3D matter power spectrum.

After projection onto $\ell$-space, the equivalent feature in the galaxy angular power spectrum can be found at
\begin{equation}
\label{eq:elleq}
\ell_{\rm eq} \equiv k_{\rm eq} D_A(0,z_i) \simeq \sqrt{\frac{2}{\omega_\gamma}} \  \int_{0}^{z_i} \di z \ \frac{ \omega_{\rm m} }{ \sqrt{\omega_{\rm m} (1+z)^3 + (h^2-\omega_{\rm m})}},
\end{equation}
where we have used $k_{\rm eq} \simeq \omega_{\rm m} \sqrt{2/\omega_\gamma}$ to obtain the last approximate equality.  Equation~(\ref{eq:elleq})  expresses a positive correlation between $\omega_{\rm m}$ and $h$  for all reasonable source galaxy redshifts~$z_i$.  Setting the source redshift explicitly to $z_i =0.8$  (i.e.,  the median redshift of the {\sc Euclid} survey) and using the power-law approximation $\ell_{\rm eq} = \omega_{\rm m}^{\mu} h^\nu$, we find $\mu \simeq 0.86$ and $\nu \simeq -0.71$ in the vicinity of
the fiducial model.  These numbers corresponds to a slope of $M \simeq 6.3$, which compares very well with $M =6.5$ from our numerical fit shown in figure~\ref{fig:2pplot}.

\paragraph{Including normalisation information}

The role of normalisation information, especially at high multipoles, can be gleaned by a close inspection of the actual expression for the galaxy angular power spectrum.  Supposing a source galaxy redshift distribution that is sharply peaked at $z_i$, the angular power spectrum can be broken down in the Limber limit into
\begin{equation}
\label{eq:clapprox}
C_{\ell, i i}^{\rm gg} \sim   \ell \  \frac{H(z_i) G^2(z_i)}{\omega_{\rm m}^2 D_A^3(0,z_i)} \ T^2_0 [k=\ell/D_A(0,z_i)] \ \mathcal{P}_{\mathcal R}[k=\ell/D_A(0,z_i)],
\end{equation}
up to a cosmology-independent constant factor.  Near scale-invariance of the primordial scalar perturbations renders $\mathcal{P}_{\mathcal R}(k)$ an almost constant factor independent of the parameters~$\omega_{\rm m}$ and~$h$ of interest.   We shall assume  $\mathcal{P}_{\mathcal R}(k)$ to be exactly constant in the following.

Note that in writing down equation~(\ref{eq:clapprox}) we have explicitly split up the transfer function~$T_\Psi(k,z)$ into a part that contains exclusively the $k$-dependence, $T_0(k)$, and one that encodes the redshift evolution, i.e, the  growth function~$G(z)$.  This splitting is in principle strictly valid only in the limit linear perturbation theory holds, and when gravitational clustering at low redshifts has no additional scale dependence (due to, e.g., hot dark matter, modified gravity, etc.).  In $\Lambda$CDM cosmologies, the (linear) growth function $G(z)$ can be written as
\begin{equation}
\label{eq:growth}
G(z) = \frac{5}{2} \, \omega_{\rm m} H(z) \int_z^{\infty} \di z' \frac{1+z'}{H^3(z')} \ ,
\end{equation}
up to a cosmology-independent conversion factor between $H_0$ and $h$.  For $T_0(k)$, the  well-known BBKS transfer function~\cite{Bardeen:1985tr},
\begin{equation}
\label{eq:bbks}
T_{\rm BBKS} \left[q = \frac{(k/{\rm Mpc})}{\omega_{\rm m}}
\right] = \frac{\ln(1+2.34 q)}{2.34 q} [1 + 3.89 q + (16.2 q)^2 + (5.47 q)^3 + (6.71 q)^4]^{-0.25},
\end{equation}
suffices as a simple and convenient approximation of $T_0(k)$.

To evaluate the degeneracy direction in the $(\omega_{\rm m}, h)$-plane, we use again  the power-law approximation $C_{\ell, i i}^{\rm gg} = \omega_{\rm m}^\mu h^\nu$, and perturb around the fiducial model.  Assuming the median source redshift $z_i =0.8$ and $\ell = \ell_{\rm max}^{{\rm g},i} \simeq 400$,  we find   $\mu \simeq 1.2$, $\nu \simeq 0.29$, and hence  a degeneracy slope of $M \simeq -22$, which matches closely $M=-21$ found in the actual fit in figure~\ref{fig:2pplot}. Interestingly, the inclusion of normalisation information now leads to a {\it negative} correlation between $\omega_{\rm m}$ and $h$, in contrast to the {\it positive} correlation we found earlier from shape information only.

\subsubsection{Shear angular power spectrum}

Shear measurements do not suffer from tracer bias effects, and therefore always contain normalisation information.  To understand the degeneracy in the $(\omega_{\rm m},h)$-plane, we again assume a sharply peaked source galaxy redshift distribution at $z_i$, and break down the shear power spectrum in the Limber limit as
\begin{equation}
\label{eq:clapprox2}
C_{\ell, i i}^{\rm ss} \sim   \ell \   \int _0^{z_i}  \di z \ \frac{G^2(z)}{H(z) D_A(0,z)} \left[ \frac{D_A(z,z_i)}{D_A(0,z_i) }\right]^2
\ T^2_0 [k=\ell/D_A(0,z)] \ \mathcal{P}_{\mathcal R}[k=\ell/D_A(0,z)],
\end{equation}
up to a cosmology-independent constant factor.  The expression can be further simplified if we assume (i) exact scale-invariance of the primordial perturbations, which reduces
$\mathcal{P}_{\mathcal R}(k)$ to an irrelevant constant, and (ii) that the gravitational lenses are all concentrated at one redshift $z=z_{\rm lens}$.  This last assumption amounts to replacing
all functional dependences on $z$ in equation~(\ref{eq:clapprox2}) with a dependence on $z_{\rm lens}$, so that no $z$-integration is required.  For $z_i  = 0.8$, the choice of $z_{\rm lens} =0.5$ should suffice.

Assuming $C_{\ell, i i}^{\rm ss} = \omega_{\rm m}^\mu h^\nu$, we find $\mu \simeq 2.8$ and $\nu \simeq -2.5$, so that $M \simeq 5.9$ in the
vicinity of the fiducial model.  These numbers have been obtained assuming $\ell \simeq 1000$, although other choices of $\ell$ yield virtually the same result for the slope $M$.
In contrast to our previous results, this value of $M$ is not a very good match to $M=9.5$ from our numerical fits in figure~\ref{fig:2pplot}.  This mismatch is most likely due to our inadequate
modelling of the nonlinear transfer function (the BBKS transfer function~(\ref{eq:bbks}) describes only the linear regime), which is important for cosmic shear at large angular multipoles.
Despite this mild  inconsistency, our result does demonstrate that the correlation between $\omega_{\rm m}$ and $h$ from shear alone is positive.  This is interesting when compared with the negative correlation preferred by the  galaxy angular power spectrum including normalisation, because na\"{\i}vely one would expect the shear and the galaxy power spectra to be measuring the same thing (at least in simple $\Lambda$CDM cosmologies).  The different normalisations and weights ${\mathcal W}_i^X$ for galaxies and shear do apparently have a strong effect on cosmological parameter estimation.

\begin{figure}[t]
\center
\includegraphics[height=.48\textwidth,angle=270]{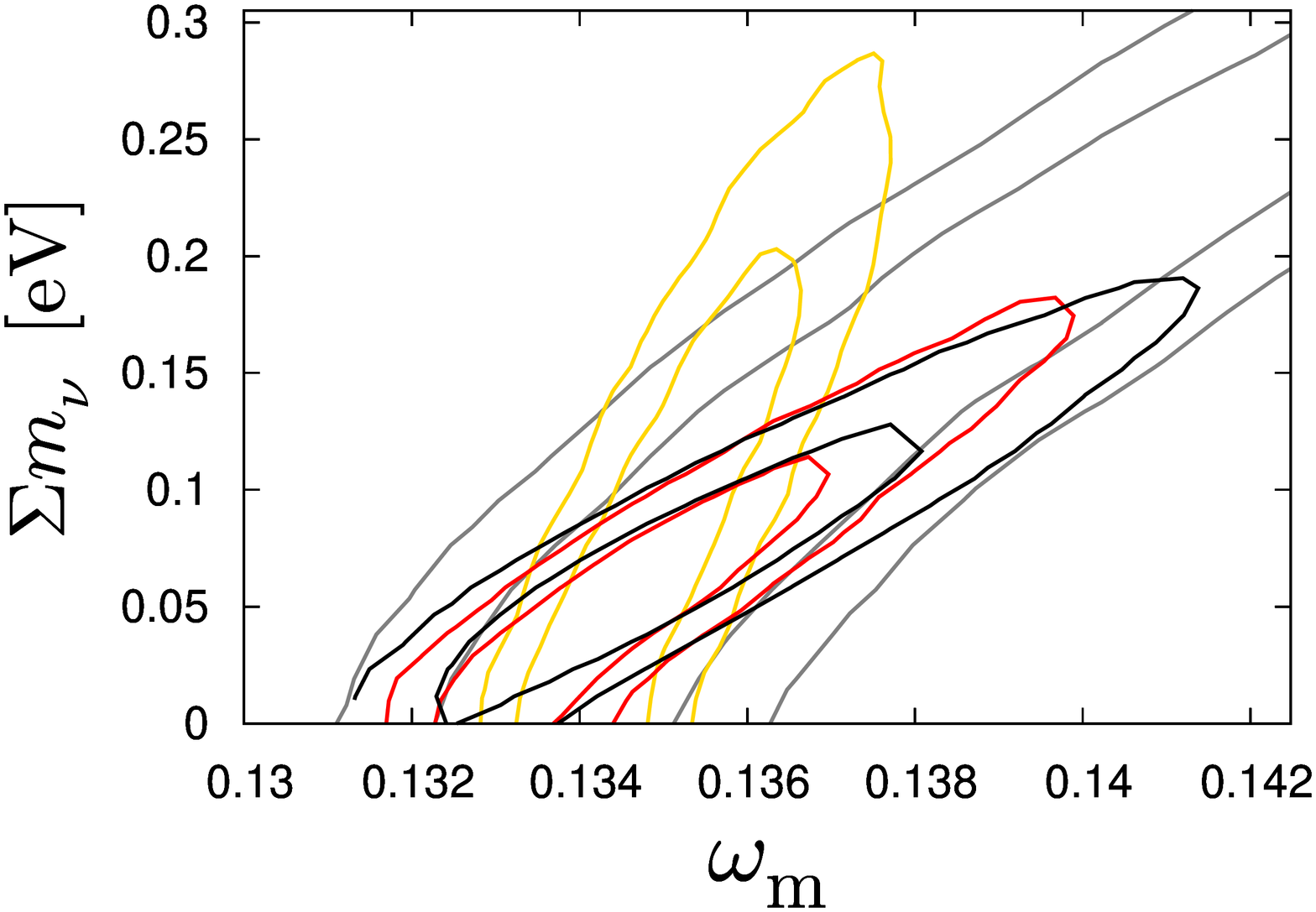}
\includegraphics[height=.48\textwidth,angle=270]{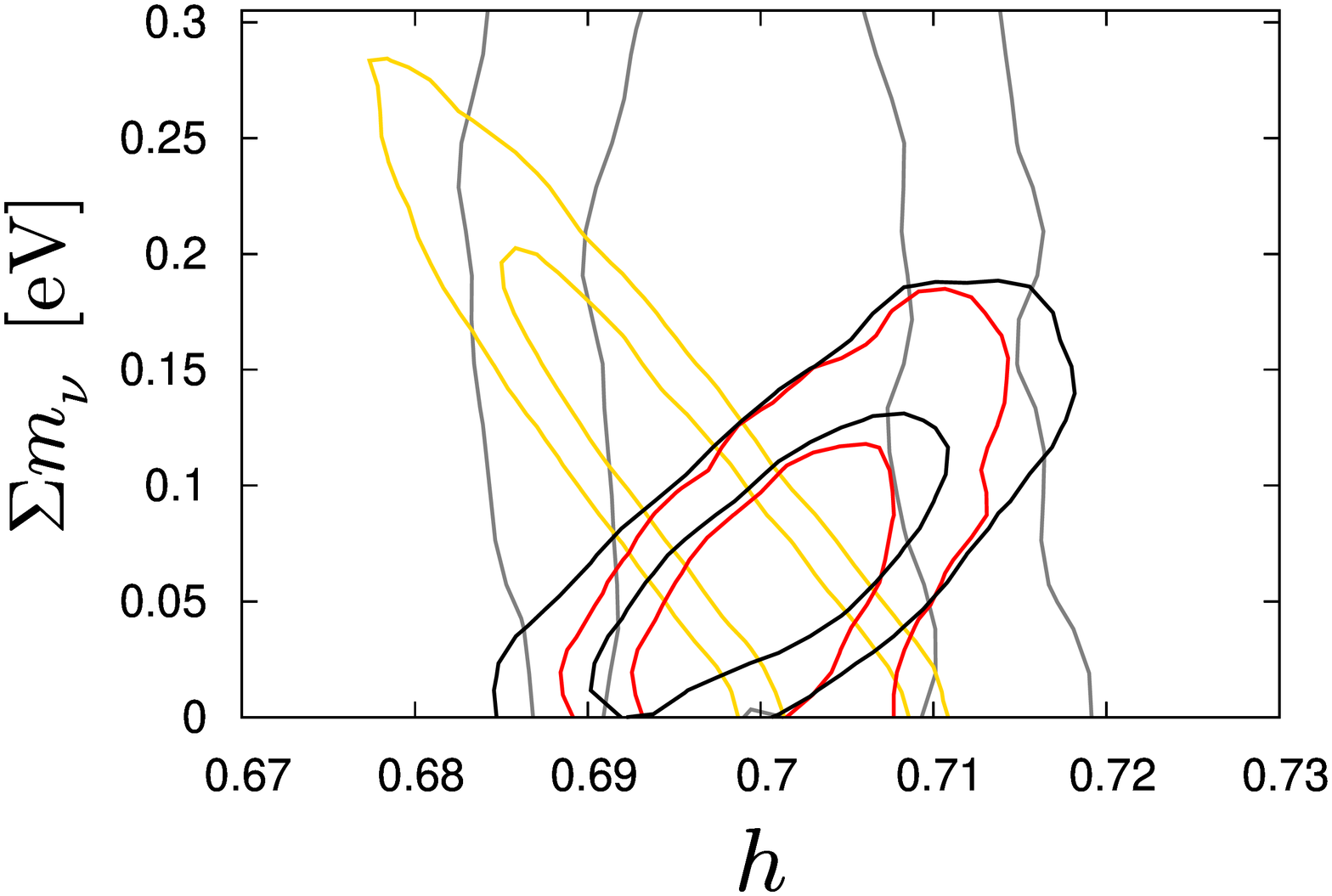}
\includegraphics[height=.48\textwidth,angle=270]{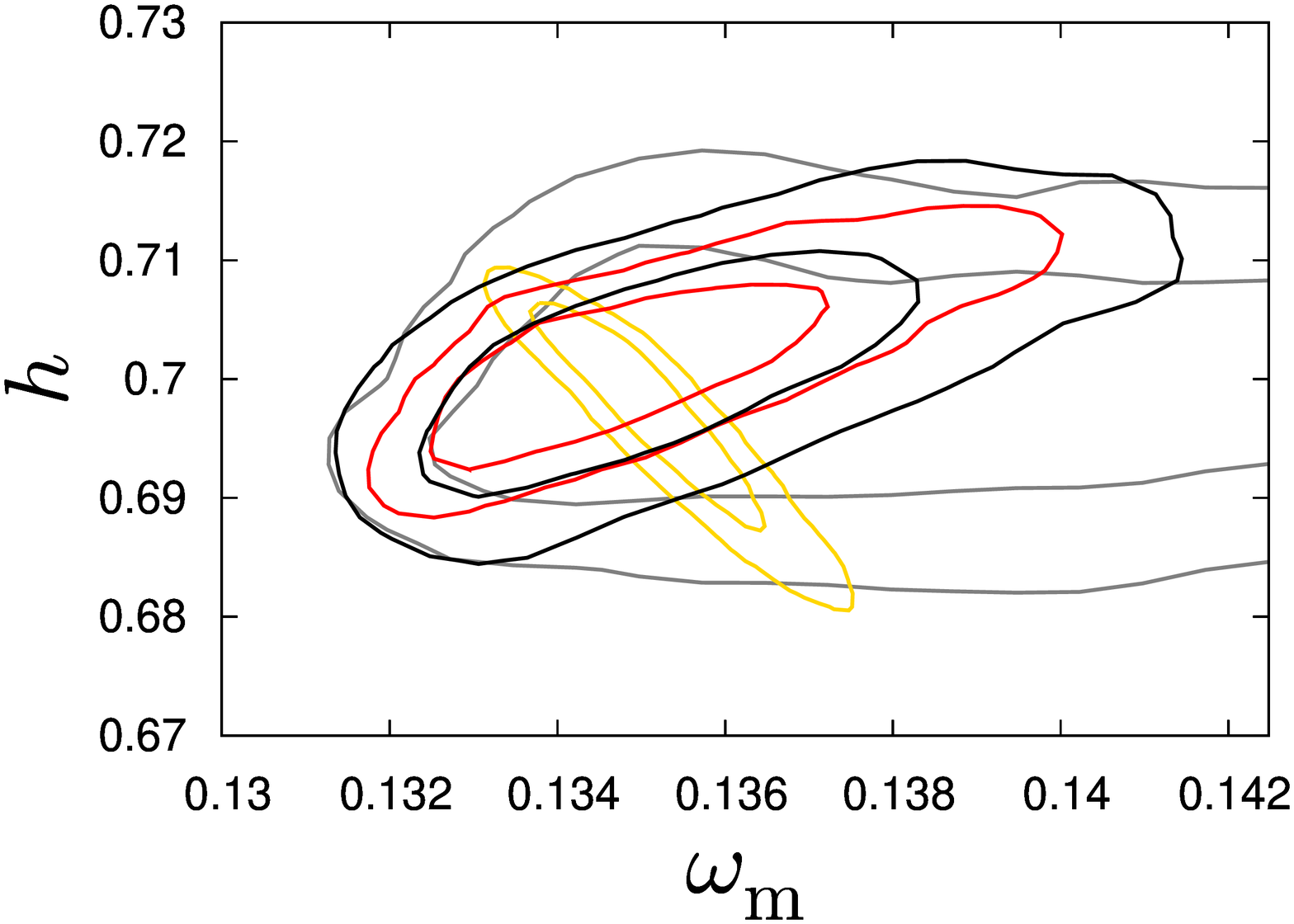}
\caption{ \label{fig:3pplots} Marginalised joint two-dimensional forecast 68\%- and 95\%-credible contours for $h$, $\omega_{\rm m}$ and $\sum m_\nu$ for the three-parameter model. The colour scheme is the same  as that in figure \ref{fig:2pplot}.}
\end{figure}

\subsection{From two to three parameters}

Adding to the fit a freely varying neutrino mass $\sum m_\nu$  complicates the picture considerably, and precludes semi-analytic degeneracy analyses such as those presented in the previous section for the two-parameter case.  One reason for this is that there are no simple approximations to the transfer function that incorporate the effects
of $\sum m_\nu$.  Another is that it is not obvious how one would project a three-dimensional degeneracy onto two dimensions,
so as to explain our three-parameter fit results in figure~\ref{fig:3pplots}, without resorting to some sort of simulation.

\begin{figure}[t]
\center
\includegraphics[width=.48\textwidth]{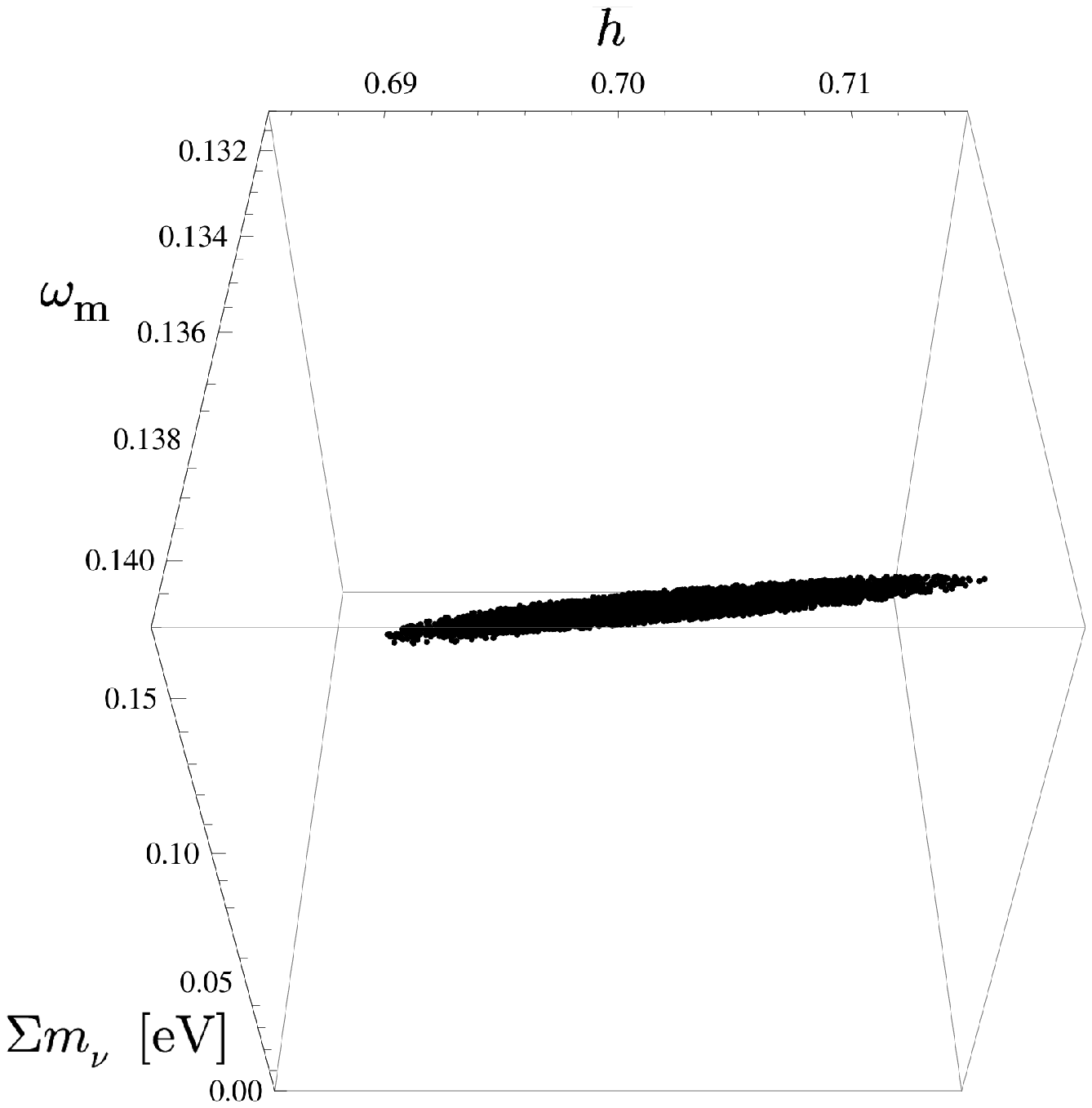}
\includegraphics[width=.48\textwidth,trim=0 -15 0 0 0]{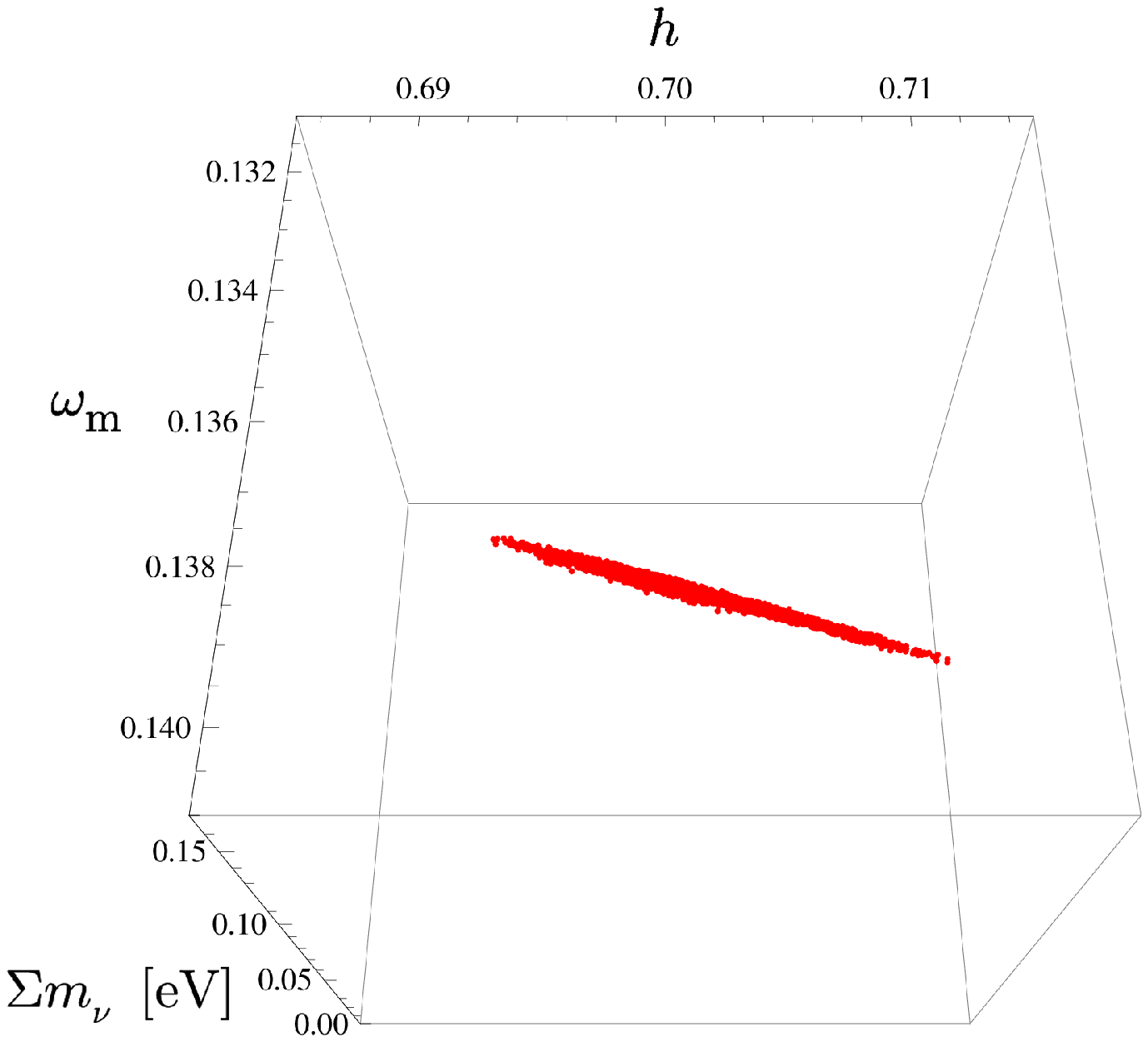}
\caption{ \label{fig:3d3pplot} Three-dimensional scatter plots of samples with $\chi_{\rm eff}^2 \leq 2$ from the Markov chains of our three-parameter fits.
 {\it Black}: Galaxy auto-spectrum only. {\it Red}: Shear auto-spectrum only.  The boxes have been rotated in the $(\omega_{\rm m},\sum m_\nu)$-plane such that all samples appear to lie on as  thin a line as possible.\label{fig:scatter}}
\end{figure}

Instead, we present an {\it a posteriori} plausibility argument.  Two plots in figure~\ref{fig:scatter} show samples with $\chi_{\rm eff}^2 \leq 2$ in $(\omega_{\rm m}, h, \sum m_\nu)$-space from the Markov chains of our three-parameter fits of the galaxy and the shear power spectra, respectively.
The galaxy power spectrum fit does not marginalise over the linear galaxy bias.
The horizontal axes in the plots are always aligned with the $h$-direction, while the boxes have been rotated around the $h$-axis
in the  $(\omega_{\rm m},\sum m_\nu)$-plane in such a way that  all samples appear to lie on as thin a line as possible.  The direction perpendicular to the line therefore corresponds the parameter direction that is the most well-constrained.

To quantify the rotation, we note that it is in fact equivalent to defining an effective parameter $\omega_{\rm eff}$ that is a linear combination of $\omega_{\rm m}$ and $\sum m_\nu$.   As shown in figure~\ref{fig:scatter}, shear and galaxy measurements require different amounts of rotation, and hence also different effective parameters $\omega_{\rm eff}$.  Phenomenologically, we find that
\begin{align}
\omega_{\rm eff, g} &= \omega_{\rm m} - 5.0 \ \omega_\nu, \\
\omega_{\rm  eff, s} &= \omega_{\rm m} - 3.0 \ \omega_\nu
\end{align}
work well for galaxies and shear respectively.  Furthermore, once these rotations have been performed, the degeneracies between $(\omega_{\rm eff,g},h)$, and
between $(\omega_{\rm eff ,s},h)$ become essentially identical to those between $(\omega_{\rm m},h)$  in the two-parameter fits  presented in figure~\ref{fig:2pplot}.
 This observation indicates that it is the effective parameters $\omega_{\rm eff,g}$ and $\omega_{\rm eff,s}$ to which galaxy and shear measurements are primarily
 sensitive, not $\omega_{\rm m}$ and $\sum m_\nu$ individually.

Looking back at the two-dimensional marginalised posteriors in figure~\ref{fig:3pplots}, we now see that the widening of the $(\omega_{\rm m},h)$-ellipses in the case of shear is simply a consequence of the measurement being sensitive to $\omega_{\rm eff,s}$ rather than to $\omega_{\rm m}$ itself, the latter of which can now take on a wider range of values because of a freely varying $\sum m_\nu$.  In the case of the galaxy power spectrum, the positive correlation we now observe between $\omega_{\rm m}$ and $h$ in the three-parameter fit can also be justified:  a negative correlation still exists between $\omega_{\rm eff, g}$ and $h$ (as it did between $\omega_{\rm m}$ and $h$ in the two-parameter fit),
but $\omega_{\rm m}$ itself can be positively correlated with $h$, so long as $\sum m_\nu$ can be adjusted accordingly.   We cannot however prove that this flipping of the degeneracy direction must occur.

\bibliographystyle{utcaps}


\providecommand{\href}[2]{#2}\begingroup\raggedright\endgroup

\end{document}